\documentclass[a4paper, 11pt]{article}
\usepackage[pdftex, a4paper]{geometry}
\usepackage{graphicx}				
		
\usepackage{cite}
\newcommand*\mcap{\mathbin{\mathpalette\mcapinn\relax}}
\newcommand*\mcapinn[2]{\vcenter{\hbox{$\mathsurround=0pt
  \ifx\displaystyle#1\textstyle\else#1\fi\bigcap$}}}

\newcommand*\mcup{\mathbin{\mathpalette\mcupinn\relax}}
\newcommand*\mcupinn[2]{\vcenter{\hbox{$\mathsurround=0pt
  \ifx\displaystyle#1\textstyle\else#1\fi\bigcup$}}}

\DeclareFontFamily{OT1}{pzc}{}
\DeclareFontShape{OT1}{pzc}{m}{it}{<-> s * [1.200] pzcmi7t}{}
\DeclareMathAlphabet{\mathpzc}{OT1}{pzc}{m}{it}

\usepackage{amsmath}
\usepackage{amsfonts}
\usepackage{dsfont}
\usepackage{mathrsfs}
\usepackage{amssymb}
\usepackage{bbm}
\usepackage{epsfig}
\usepackage[english]{babel}
\usepackage[all]{xy}
\usepackage{subfigure}
\usepackage{color}
\def\T{^{\rm\tiny T}}

\newtheorem{theorem}{Theorem}
\newtheorem{definition}{Definition}

\newtheorem{lemma}{Lemma}
\newtheorem{remark}{Remark}

\title{\bf Network  Flows that Solve Linear   Equations}

\author{Guodong Shi, Brian D. O. Anderson and Uwe Helmke\thanks{A brief version of the current manuscript  was presented at  the American Control Conference in Boston, July 2016 \cite{acc2016}.
}\thanks{G. Shi is with the Research School of Engineering, College of Engineering and Computer Science, The Australian National University, Canberra 0200, Australia. E-mail: guodong.shi@anu.edu.au.}
\thanks{B. D. O. Anderson is with the National ICT Australia (NICTA) and the Research School of Engineering, College of Engineering and Computer Science, The Australian National University, Canberra 0200, Australia. E-mail: brian.anderson@anu.edu.au.}
\thanks{U. Helmke is deceased but was  with  Institute of Mathematics, University of W\"{u}rzburg, 97074 W\"{u}rzburg, Germany. }}
\date{}

\begin{document}

\maketitle

\begin{abstract}
We study  distributed network flows as  solvers  in continuous time for the linear algebraic equation $\mathbf{z}=\mathbf{H}\mathbf{y}$. Each node $i$ has access to a row $\mathbf{h}_i\T$ of the matrix $\mathbf{H}$ and  the corresponding entry $z_i$ in the vector $\mathbf{z}$. The first ``consensus + projection" flow under investigation  consists of two terms, one from standard consensus dynamics and the other contributing  to projection onto each affine subspace specified by the $\mathbf{h}_i$ and $z_i$. The second ``projection consensus" flow on the other hand simply replaces the relative state feedback in consensus dynamics with projected relative state feedback.
Without  dwell-time assumption on switching graphs, we prove that all node states converge to a common solution of the linear algebraic equation, if there is any. The convergence is global for the ``consensus + projection" flow while local for the ``projection consensus" flow in the sense that the initial values must lie on the affine subspaces.  If the linear equation has no exact solutions, we show that the node states can converge to a ball around the least squares solution whose radius can be made arbitrarily small through selecting a sufficiently large gain for the ``consensus + projection" flow for a fixed bidirectional graph. Semi-global convergence to approximate least squares solutions is also demonstrated  for switching balanced directed graphs under suitable conditions. It is also shown that the ``projection consensus" flow drives the average of the node states to the least squares solution with a complete graph. Numerical examples are provided as illustrations of the established results.
\end{abstract}
\section{Introduction}

In the past decade, distributed consensus algorithms have attracted a significant amount of research attention \cite{jad03, xiao04,  saber04,  ren05}, due to their wide applications in distributed control and estimation \cite{martinez07,kar12}, distributed signal processing \cite{Rabbat2010}, and distributed optimization methods \cite{rabbat2004,nedic09}. The basic idea is that a network of interconnected nodes with different initial values can reach a common value, namely an  agreement or a consensus, via local information exchange as long as the communication graph is well connected. The consensus value is informative in the sense that it can depend on all nodes' initial states  even if the value is not the exact average \cite{jackson}; the consensus processes are robust against random or deterministic switching of network interactions as well as against noises \cite{moreau05, lin07, jad08, shisiam}.

As a generalized  notion, constrained consensus seeks to reach state agreement  in the intersection of a number of convex sets, where each set serves  as the   supporting state space for a particular node \cite{nedic10}. It was shown that with doubly stochastic arc weights, a projected consensus algorithm,  where each node iteratively projects a weighted average of its neighbor's values onto its supporting set,  can guarantee convergence to constrained consensus when the intersection set is bounded and contains an interior point \cite{nedic10}. This idea was then extended to the continuous-time case under which the graph connectivity and arc weight conditions can be relaxed with convergence still being guaranteed \cite{shitac}. The projected consensus algorithm was shown to be convergent  even under randomized projections \cite{shiauto}.  In fact, those developments are closely related to a class of so-called alternating projection algorithms,  which was first studied by von Neumann in the 1940s \cite{jvn49} and then extensively investigated across the past many decades \cite{aron50, gubin1967,deut83,B-B-SIAM}.

A related but different   intriguing problem  lies in how to develop distributed algorithms that solve a linear algebraic equation $\mathbf{z}=\mathbf{H}\mathbf{y}, \ \mathbf{H}\in\mathbb{R}^{N\times m},\mathbf{z}\in\mathbb{R}^N $ with respect to variable $\mathbf{y}\in \mathbb{R}^m$, to which tremendous efforts have been devoted with many algorithms developed under different assumptions of information exchanges among the nodes \cite{anderson97, mehmood05, lu09-1,lu09-2,liu13,mou13,Morse-TAC15, asuman14, jadbabaie15, brian15}. Naturally, we can assume that a node $i$ has access to a row vector, $\mathbf{h}_i\T$  of $\mathbf{H}$ as well as the corresponding entry,  $z_i$ in $\mathbf{z}$. Each node will then be able to determine with straightforward calculations an affine subspace whose elements satisfy the equation $\mathbf{h}_i\T \mathbf{y}=z_i$. This means that, as long as the original equation $\mathbf{z}=\mathbf{H}\mathbf{y}$ has a nonempty solution set, solving the equation would be equivalent to finding a point in the intersection of all the affine subspaces and therefore  can be put into the constrained consensus framework. If however the linear equation has no exact solution and  its least squares solutions are of interest, we ended up with a convex optimization problem with quadratic cost and linear constraints. Many distributed optimization algorithms, such as \cite{nedic09,nedic10,elia,jmf,nedic11,Rabbat2012,lu12,cotes14,jakovetic14}, developed for much more complex models, can therefore be directly applied. It is worth emphasizing that the above ideas of distributed consensus and optimization were traced back to the seminal work of Bertsekas and J. Tsitsiklis \cite{tsi,tsibook}.

In this paper, we study two distributed network flows as distributed solvers for such linear algebraic equations in continuous time. The first so-called ``consensus + projection" flow consists of two additive  terms, one from standard consensus dynamics and the other contributing to projection onto each affine subspace. The second ``projection consensus" flow on the other hand simply replaces the relative state feedback in consensus dynamics with projected relative state feedback. Essentially only relative state information is exchanged among the nodes for both of the two flows, which justifies their full {\it distributedness}.  To study the asymptotic behaviours of the two distributed flows with respect to the solutions of the linear equation, new challenges arise in the loss of intersection boundedness and interior points for the exact solution case as well as in the complete loss of intersection sets for the least squares solution case. As a result, the analysis cannot be simply mapped back to the studies in \cite{nedic10,shitac}.

The contributions of the current paper are summarized as follows.

\begin{itemize}
\item Under mild conditions on the communication graphs (without requiring a  dwell-time for switching graphs and  only  requiring a  positively lower bound on the integral of arc weights over certain time intervals), we prove that all node states asymptotically converge to a common solution of the linear algebraic equation for the two flows, if there is any. The convergence is global for the ``consensus + projection" flow, and local for the ``projection consensus" flow in the sense that the initial values must be put into the affine subspaces  (which is a very minor restriction indeed). We manage to characterize the node limits for balanced or fixed graphs.

\item If the linear equation has no exact solutions, we show that the node states can be forced to converge to a ball of fixed but arbitrarily small radius surround the  least squares solution by taking  the gain of the consensus dynamics to be sufficiently large for ``consensus + projection" flow under fixed and undirected graphs. Semi-global convergence to approximate least squares solutions is established  for switching balanced directed graphs under suitable conditions.  A minor, but more explicit result arises where it is also shown that the ``projection consensus" flow drives the average of the node states to the least squares solution with complete communication graphs.
\end{itemize}

These results rely on our development of new technique of independent interest for treating the interplay between consensus dynamics and the projections onto affine subspaces in the absence of  intersection boundedness and interior point assumptions. All the convergence results  can be sharpened to provide exponential convergence with suitable conditions on the switching graph signals.

The remainder of this paper is organized as follows. Section \ref{Sec:Problem} introduces the network model, presents the distributed flows under consideration, and defines the problem of interest. Section \ref{Sec:exact} discusses the scenario where the linear equation has at least one solution. Section \ref{Sec:least-squares} then turns to the least squares case where the linear equation has no solution at all. Finally, Section \ref{Sec:numerical} presents a few numerical examples illustrating the established  results and Section \ref{Sec:conclusions} concludes the paper with a few remarks.

\subsection*{Notation and Terminology}

A directed graph (digraph) is an ordered pair of two sets denoted by $\mathrm {G} =(\mathrm{V}, \mathrm{E} )$ \cite{god}. Here $\mathrm{V}=\{1,\dots,N\}$  is a finite set of  vertices (nodes). Each element in $\mathrm{E}$ is an ordered pair of two distinct  nodes in $\mathrm {V}$, called an arc.  A  directed path in $\mathrm {G}$ with length $k$ from $v_1$ to $v_{k+1}$ is a  sequence of distinct nodes, $v_1v_2\dots v_{k+1}$, such that  $(v_m, v_{m+1}) \in \mathrm{E}$, for all $m=1,\dots,k$. A digraph $\mathrm{G}$ is termed {\it strongly connected} if for any two distinct nodes $i,j\in\mathrm{V}$, there is a  path from $i$ to $j$. A digraph is called {\it bidirectional} when $(i,j)\in\mathrm{E}$ if and only if $(j,i)\in\mathrm{E}$ for all $i$ and $j$. A strongly connected bidirectional digraph  is  simply called {\it connected}. All vectors are column vectors and denoted by bold, lower case letters, i.e., $\mathbf{a},\mathbf{b},\mathbf{c}$,  etc.; matrices are denoted with bold, upper case letters, i.e.,  $\mathbf{A},\mathbf{B},\mathbf{C}$,  etc.;  sets are denoted with $\mathcal{A},\mathcal{B},\mathcal{C}$, etc. Depending on the argument, $|\cdot|$ stands for the absolute value of a real number or the cardinality of a set.  The Euclidean inner product between two vectors $\mathbf{a}$ and $\mathbf{b}$ in $\mathbb{R}^m$ is denoted as $\langle \mathbf{a}, \mathbf{b}\rangle$, and sometimes  simply  $\mathbf{a}\T \mathbf{b}$.  The Euclidean norm of a vector is denoted as $\|\cdot\|$.


\section{Problem Definition}\label{Sec:Problem}

\subsection{Linear Equations}
Consider the following  linear algebraic equation:
\begin{align}\label{LinearEquation}
\mathbf{z}=\mathbf{H} \mathbf{y}
\end{align}
with respect to variable $\mathbf{y}\in\mathbb{R}^m$, where $\mathbf{H}\in\mathbb{R}^{N\times m}$ and $\mathbf{z}\in\mathbb{R}^N$. We know from the basics of linear algebra that overall there are three cases.

\begin{itemize}
\item[(I)] There exists a unique solution satisfying  Eq. (\ref{LinearEquation}): ${\rm rank}(\mathbf{H})=m$ and $\mathbf{z}\in {\rm span}(\mathbf{H})$.

\item[(II)] There is an infinite set of  solutions  satisfying  Eq. (\ref{LinearEquation}): ${\rm rank}(\mathbf{H})<m$ and $\mathbf{z}\in {\rm span}(\mathbf{H})$.

\item[(III)] There exists no solution   satisfying  Eq. (\ref{LinearEquation}): $\mathbf{z}\notin {\rm span}(\mathbf{H})$.
\end{itemize}

We denote
$$
\mathbf{H}= \begin{pmatrix}
  \mathbf{h}_1\T \\
 \mathbf{h}_2\T \\
  \vdots  \\
\mathbf{h}_N\T
 \end{pmatrix}, \quad \mathbf{z}= \begin{pmatrix}
  z_1 \\
 z_2 \\
  \vdots  \\
z_N
 \end{pmatrix}
$$
with $\mathbf{h}_i\T$ being the $i$-th row vector of $\mathbf{H}$. For the ease of presentation and with inessential loss of generality we assume throughout the rest of the paper that
$$
\big\|\mathbf{h}_i\big\|=1,\ i=1,\dots,N.
$$ Introduce
\begin{align*}
\mathcal{A}_i:=\Big\{\mathbf{y}:\ \mathbf{h}_i\T \mathbf{y}=z_i \Big\}
\end{align*}
for each $i=1,\dots,N$, which is an affine subspace.  It is then clear that Case (I) is equivalent to the condition that $\mathcal{A}:= \mcap_{i=1}^N\mathcal{A}_i$ is a singleton, and that Case (II) is equivalent to the condition  that $\mathcal{A}:= \mcap_{i=1}^N\mathcal{A}_i$ is an affine space with a nontrivial dimension. For Case (III), a least squaress solution of (\ref{LinearEquation}) can be defined via the following  optimization problem:
\begin{align}
\min_{\mathbf{y}\in\mathbb{R}^m} \big\|\mathbf{z}-\mathbf{H}\mathbf{y}\big\|^2,
\end{align}
which yields a unique solution $\mathbf{y}^\star=(\mathbf{H}\T\mathbf{H})^{-1}\mathbf{H}\mathbf{z}$ if ${\rm rank}(\mathbf{H})=m$.

Consider a network with nodes  indexed in the set $\mathrm{V}=\big\{1,\dots,N\big\}$.  Each node $i$ has access to the value of $\mathbf{h}_i$ and ${z}_i$ without the knowledge of $\mathbf{h}_j$ or ${z}_j$ from other nodes. Each node $i$  holds a state $\mathbf{x}_i(t) \in\mathbb{R}^m$ and exchanges this state information with other neighbor nodes, these being determined by the edges of the graph of the network.
We are interested in distributed flows for the $\mathbf{x}_i(t)$ that asymptotically solve the equation (\ref{LinearEquation}), i.e., $\mathbf{x}_i(t)$ approaches some solution of  (\ref{LinearEquation}) as $t$ grows.
\subsection{Network Communication Structures}
Let $\Theta$ denote the set of all directed graphs associated with node set $\mathrm{V}$. Node interactions are described by a  signal $\sigma(\cdot): \mathbb{R}^{\geq 0}\mapsto \Theta$.
The digraph that $\sigma(\cdot)$ defines at time $t$ is denoted as $\mathrm{G}_{\sigma(t)}=\big(\mathrm{V}, \mathrm{E}_{\sigma(t)}\big)$, where $\mathrm{E}_{\sigma(t)}$ is the set of arcs. The neighbor set of node $i$ at time $t$, denoted $\mathrm{N}_i(t)$, is given by
\begin{align*}
\mathrm{N}_i(t):= \Big\{j:\ (j,i)\in\mathrm{E}_{\sigma(t)} \Big\}.
\end{align*}
This is to say, at any given time $t$, node $i$ can only receive information from the nodes in the set $\mathrm{N}_i(t)$.

Let $\mathbb{R}^{\geq 0}$ and $\mathbb{R}^+$ be the sets of nonnegative and positive real numbers, respectively. Associated with each ordered pair $(j,i)$ there is a function $a_{ij}(\cdot): \mathbb{R}^{\geq 0} \rightarrow \mathbb{R}^+$ representing the weight of the possible connection $(j,i)$. We impose the following assumption, which will be adopted throughout the paper without specific  further mention.

\medskip

\noindent {\bf Weights Assumption.} The function $a_{ij}(\cdot)$ is continues except for at most a set with measure zero over $\mathbb{R}^{\geq 0}$ for all $i,j\in\mathrm{V}$; there exists $a^\ast>0$ such that $ a_{ij}(t)\leq a^\ast$ for all $t\in \mathbb{R}^{\geq 0}$ and all $i,j\in\mathrm{V}$.

\medskip

Denote $\mathbb{I}_{(j,i)\in \mathrm{E}_{\sigma(t)}}$ as an indicator function for all $i\neq j\in \mathrm{V}$, where  $\mathbb{I}_{(j,i)\in \mathrm{E}_{\sigma(t)}}=1$ if $(j,i)\in \mathrm{E}_{\sigma(t)}$ and $\mathbb{I}_{(j,i)\in \mathrm{E}_{\sigma(t)}}=0$ otherwise.  We impose the following definition on the connectivity of the network communication structures.

\begin{definition}
\begin{itemize}
\item[(i)] An arc $(j,i)$ is said to be a $\delta$-arc of $\mathrm{G}_{\sigma(t)}$ for the
time interval $[t_1,t_2)$  if $$
\int_{t_1}^{t_2}a_{ij}(t) \mathbb{I}_{(j,i)\in \mathrm{E}_{\sigma(t)}}dt\geq \delta.
$$

\item[(ii)] $\mathrm{G}_{\sigma(t)}$ is $\delta$-uniformly jointly strongly connected ($\delta$-$\mathsf{UJSC}$) if there exists $T>0$ such that  the  $\delta$-arcs of  $\mathrm{G}_{\sigma(t)}$ on time interval $[s,s+T)$ form a  strongly connected digraph for all $s\geq 0$;

\item[(iii)] $\mathrm{G}_{\sigma(t)}$ is $\delta$-bidirectionally infinitely  jointly  connected ($\delta$-$\mathsf{BIJC}$) if   $\mathrm{G}_{\sigma(t)}$ is bidirectional for all $t\geq 0$ and the  $\delta$-arcs of  $\mathrm{G}_{\sigma(t)}$ on time interval $[s,\infty)$ form a connected graph for all $s\geq 0$.
    \end{itemize}
\end{definition}
\subsection{Distributed Flows}
Define the mapping $\mathpzc{P}_{\mathcal{A}_i}: \mathbb{R}^m \rightarrow \mathbb{R}^m$ as the projection onto the affine subspace $\mathcal{A}_i$. Let $K>0$ be a given constant. We consider the following continuous-time network flows:

\medskip

\noindent[{\bf ``Consensus + Projection" Flow}]:
\begin{align}\label{1}
 \dot{\mathbf{x}}_i =K\Big( \sum_{j\in \mathrm{N}_i(t)}a_{ij}(t)\big(\mathbf{x}_j-\mathbf{x}_i\big) \Big)+ \mathpzc{P}_{\mathcal{A}_i}(\mathbf{x}_i)-\mathbf{x}_i,\ \  i\in\mathrm{V};
\end{align}
\noindent[{\bf ``Projection Consensus" Flow}]:
\begin{align}\label{2}
 \dot{\mathbf{x}}_i =  \sum_{j\in \mathrm{N}_i(t)}a_{ij}(t)\big(\mathpzc{P}_{\mathcal{A}_i}(\mathbf{x}_j)-\mathpzc{P}_{\mathcal{A}_i}(\mathbf{x}_i)\big),\ \  i\in\mathrm{V}.
\end{align}

\begin{figure*}
\begin{minipage}[t]{0.5\linewidth}
\centering
\includegraphics[width=2.4in]{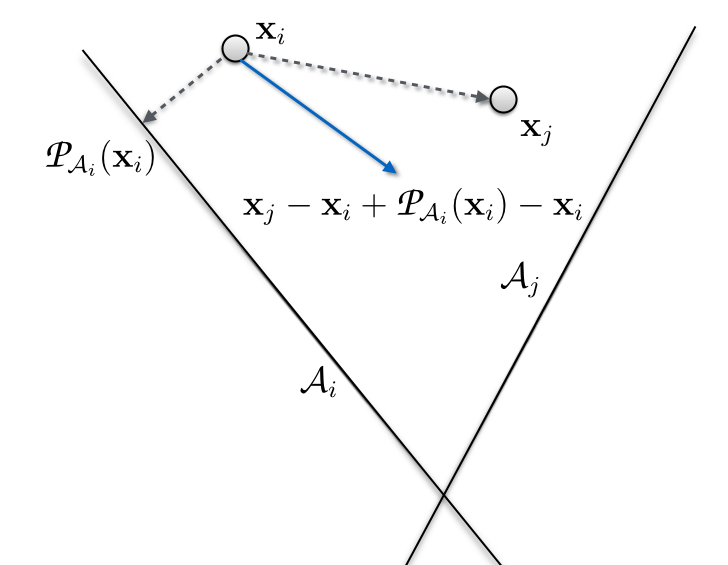}
\end{minipage}%
\begin{minipage}[t]{0.5\linewidth}
\centering
\includegraphics[width=2.4in]{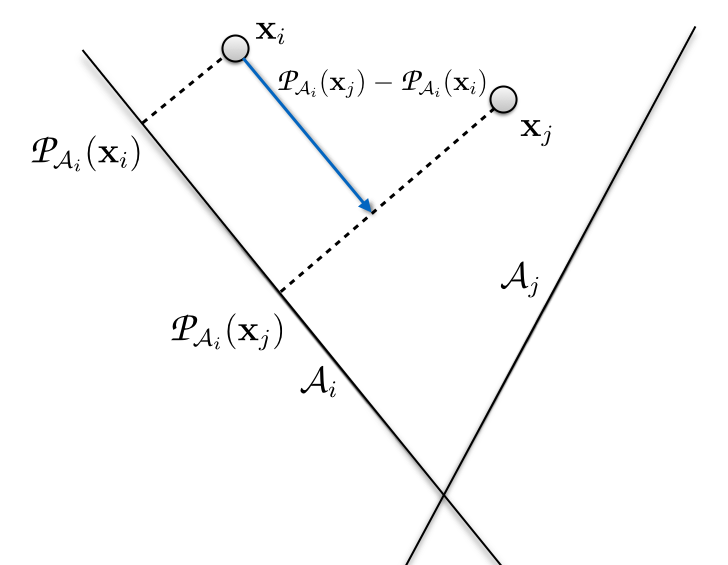}
\end{minipage}
\caption{An illustration of the ``consensus + projection" flow (left) and the ``projection consensus" flow (right). The blue arrows mark the vector of $\dot{\mathbf{x}}_i$ for the two flows, respectively.}
\label{fig:twoflows}
\end{figure*}

The first term of (\ref{1}) is a standard  consensus flow \cite{jad03}, while along
\begin{align}\label{r1}
 \dot{\mathbf{x}}_i= \mathpzc{P}_{\mathcal{A}_i}(\mathbf{x}_i)-\mathbf{x}_i,
\end{align}
$x_i(t)$ will be asymptotically projected onto ${\mathcal{A}_i}$ since (\ref{r1}) is a gradient flow with $\mathpzc{P}_{\mathcal{A}_i}(v)-v =- \nabla \|v\|_{\mathcal{A}_i}^2/2$, where $\nabla \|v\|_{\mathcal{A}_i}^2$ is a  $C^\infty$ convex function. The flow (\ref{2}) simply replaces the relative state in standard consensus flow with relative projective state.

Notice that a particular equilibrium point of these equations is given by  $\mathbf{x}_1=\mathbf{x}_2=\cdots=\mathbf{x}_N=\mathbf{y}$, where $\mathbf{y}$ is a solution of Eq. (\ref{LinearEquation}). The aim of the two flows is to ensure that the $\mathbf{x}_i(t)$ asymptotically tend to a solution of Eq. (\ref{LinearEquation}). Note that, the projection consensus flow (\ref{2}) can be rewritten as\footnote{A rigorous treatment will be given later in Lemma \ref{lem:subspace-affine-projection}.}
\begin{align*}
\dot{\mathbf{x}}_i =  \mathpzc{P}_{\mathcal{A}_i}\Big( \sum_{j\in \mathrm{N}_i(t)}a_{ij}(t)(\mathbf{x}_j-\mathbf{x}_i)\Big) -  \sum_{j\in \mathrm{N}_i(t)}a_{ij}(t)\cdot\mathpzc{P}_{\mathcal{A}_i}(0).
\end{align*}
Therefore, in both of the ``consensus + projection" and the  ``projection consensus" flows, a node $i$ essentially only receives the information
$$
 \sum_{j\in \mathrm{N}_i(t)}a_{ij}(t)\big(\mathbf{x}_j(t)-\mathbf{x}_i(t)\big)
$$
from its  neighbors, and the flows can then be utilized in addition with the $\mathbf{h}_i$ and $z_i$ it holds. In this way,   the  flows (\ref{1}) and (\ref{2}) are {\it distributed}.

Without loss of generality we assume the initial time is $t_0=0$. We denote by  $\mathbf{x}(t)=(\mathbf{x}_1\T(t) \cdots \mathbf{x}_N\T(t))\T$ a trajectory along any of the two flows.

\subsection{Discussions}

\subsubsection{Geometries}
 The two network flows are intrinsically different in their geometries. In fact, the ``consensus + projection" flow  is a special case of the optimal consensus flow proposed in \cite{shitac} consisting of two parts, a consensus part and another projection part. The ``projection consensus" flow, first  proposed and studied in \cite{brian15} for fixed bidirectional graphs,  is the continuous-time analogue of the projected consensus algorithm   proposed in \cite{nedic10}. Because it is a gradient descent on a Riemannian manifold if the communication graph is undirected and fixed, there is guaranteed convergence to an equilibrium point.  The two flows are closely related to the alternating projection algorithms,  first studied by von Neumann in the 1940s \cite{jvn49}. We refer to  \cite{B-B-SIAM} for a thorough survey on the developments of alternating projection algorithms. We illustrate the intuitive  difference between the two flows in Figure \ref{fig:twoflows}.

\subsubsection{Relation with Previous Work}
The dynamics described in systems (\ref{1}) and (\ref{2}) are linear, in contrast to the nonlinear dynamics associated with the general convex sets  studied in \cite{nedic10,shitac}. However, we would like to emphasize that new challenges arise with the   systems (\ref{1}) and (\ref{2}) compared to the work of \cite{nedic10,shitac}. First of all, for both of the Cases (I) and (II), $\mathcal{A}:= \mcap_{i=1}^N\mathcal{A}_i$ contains no interior point. This interior point condition  however is essential to the  results   in   \cite{nedic10}.
Next, for Case (II) where  $\mathcal{A}:= \mcap_{i=1}^N\mathcal{A}_i$ is an affine space with a nontrivial dimension, the boundedness condition for the intersection of the convex sets no longer holds, which plays a key role in the analysis of  \cite{nedic10,shitac}. Finally, for Case (III),  $\mathcal{A}:= \mcap_{i=1}^N\mathcal{A}_i$ becomes an empty set. The least squares solution case then completely falls out from the discussions of constrained consensus in \cite{shitac,nedic10}.

\subsubsection{Grouping of Rows}

We have assumed hitherto that each node $i$ only has access to the value of $\mathbf{h}_i$ and ${z}_i$ from the equation (\ref{LinearEquation}) and therefore there are a total of $N$ nodes. Alternatively and as a generalization, there can be $n\leq N$ nodes with node $i$ holding $n_i\geq 1$ rows, denoted $(\mathbf{h}_{i(1)}\ {z}_{i(1)}),\dots,(\mathbf{h}_{i({n_i})}\ {z}_{i({n_i})})$ of $(\mathbf{H}\ \mathbf{z})$ with $i(k)\in \{1,\dots, Z\}$. In this case, we can revise the definition of $\mathcal{A}_i$ to
\begin{align*}
\mathcal{A}_i:=\Big\{\mathbf{y}:\ \mathbf{h}_{i(k)}\T \mathbf{y}=z_{i(k)},\ k=1,\dots,n_i \Big\},
\end{align*}
which is nonempty if (\ref{LinearEquation}) has at least one solution. Then the two flows (\ref{1}) and (\ref{2}) can be defined in the same manner for the $n$ nodes. Of course, if $n_i$ is large, finding an initial condition consistent with $\mathcal{A}_i$ becomes computationally more burdensome.

Let\footnote{Note that, it is not necessary to require the $\big\{ i(1), \dots, i({n_i})\big\}$ to be disjoint.} $$
\mcup_{i=1}^n \big\{ i(1), \dots, i({n_i})\big\}=\big\{ 1, \dots, Z\big\}.
$$
Note  that,  the $\mathcal{A}_i$ are still affine subspaces,  while solving (\ref{LinearEquation}) exactly  continues to be equivalent to finding a point in $\mcap_{i=1}^n \mathcal{A}_i$. Consequently, all our results for Cases (I) and (II) apply also to this new setting with row grouping.

\section{Exact Solutions}\label{Sec:exact}
In this section, we show how the two distributed flows  asymptotically  solve the equation (\ref{LinearEquation}) under quite general conditions for Cases (I) and (II).
\subsection{Singleton  Solution Set}
We first focus on the case when Eq. (\ref{LinearEquation}) admits a unique solution $\mathbf{y}_\ast$, or equivalently,  $$
\mathcal{A}:= \mcap_{i=1}^N\mathcal{A}_i=\{\mathbf{y}_\ast\}
$$ is a singleton. For the ``consensus + projection" flow (\ref{1}), the following theorem holds.


\begin{theorem}\label{thm1}
Let (I) hold. Then  along  the ``consensus + projection" flow (\ref{1}), there holds $$
\lim_{t\to \infty} \mathbf{x}_i(t)=\mathbf{y}_\ast,\ \forall i\in\mathrm{V}$$ for all initial values if $\mathrm{G}_{\sigma(t)}$ is either  $\delta$-$\mathsf{UJSC}$ or $\delta$-$\mathsf{BIJC}$.
\end{theorem}

The ``projection consensus" flow (\ref{2}), however, can only guarantee local convergence for a particular set of initial values. The following theorem holds.
\begin{theorem}\label{thm2}
Let (I) hold. Suppose $\mathbf{x}_i(0)\in \mathcal{A}_i$ for all $i$. Then  along  the  ``projection consensus" flow (\ref{2}), there holds
$$
\lim_{t\to \infty} \mathbf{x}_i(t)=\mathbf{y}_\ast,\ \forall i\in\mathrm{V}
$$ if $\mathrm{G}_{\sigma(t)}$ is either  $\delta$-$\mathsf{UJSC}$ or $\delta$-$\mathsf{BIJC}$.
\end{theorem}

\begin{remark}
Convergence along the ``projection consensus" flow relies on specific  initial values  due to the existence of equilibriums other than the desired consensus states within the set $\mathcal{A}$: if $\mathbf{x}_i(0)$ are all equal, then obviously they will stay there for ever along the flow (\ref{2}). It was suggested in \cite{brian15} that one can add another term in the ``projection consensus" flow and arrive at
\begin{align}\label{100}
 \dot{\mathbf{x}}_i =  \sum_{j\in \mathrm{N}_i(t)}a_{ij}(t)\big(\mathpzc{P}_{\mathcal{A}_i}(\mathbf{x}_j)-\mathpzc{P}_{\mathcal{A}_i}(\mathbf{x}_i)\big) + \mathpzc{P}_{\mathcal{A}_i}(\mathbf{x}_i)-\mathbf{x}_i
\end{align}
for $i\in\mathrm{V}$, then convergence will be global under (\ref{100}). We would like to point out that (\ref{100}) has a similar structure  as the ``consensus + projection" flow with the consensus dynamics being replaced by projection consensus.  In fact, our analysis of the ``consensus + projection" flow developed  in  this paper can be refined and generalized to  the flow (\ref{100}) and then leads to  results under the same graph conditions for both exact and least squares solutions.
\end{remark}

\subsection{Infinite  Solution Set}
We now turn to the scenario when Eq. (\ref{LinearEquation}) has an infinite set of  solutions, i.e., $\mathcal{A}:= \mcap_{i=1}^N\mathcal{A}_i$ is an affine space with a nontrivial dimension. We note that in this case $\mathcal{A}$ is no longer a bounded set; nor does it contain interior points. This is in contrast to the situation studied in \cite{nedic10,shitac}.

For the ``consensus + projection" flow (\ref{1}), we present the following result.

\begin{theorem}\label{thm3}
Let (II) hold. Then   along  the ``consensus + projection" flow (\ref{1}) and for any initial value $\mathbf{x}(0)$, there exists  $\mathbf{y}^\flat(\mathbf{x}(0))$, which is a solution of (\ref{LinearEquation}), such that   $$
\lim_{t\to \infty} \mathbf{x}_i(t)=\mathbf{y}^\flat(\mathbf{x}(0)),\  \forall i\in\mathrm{V}
$$ if $\mathrm{G}_{\sigma(t)}$ is either  $\delta$-$\mathsf{UJSC}$ or $\delta$-$\mathsf{BIJC}$.
\end{theorem}

For the ``projection consensus" flow, convergence relies on restricted initial nodes states.
\begin{theorem}\label{thm4}
Let (II) hold. Then   along  the `projection consensus" flow  (\ref{2}) and for any initial value $\mathbf{x}(0)$ with $\mathbf{x}_i(0)\in \mathcal{A}_i$ for all $i$, there exists  $\mathbf{y}^\flat(\mathbf{x}(0))$, which is a solution of (\ref{LinearEquation}), such that   $$
\lim_{t\to \infty} \mathbf{x}_i(t)=\mathbf{y}^\flat(\mathbf{x}(0)),\  \forall i\in\mathrm{V}
$$ if $\mathrm{G}_{\sigma(t)}$ is either  $\delta$-$\mathsf{UJSC}$ or $\delta$-$\mathsf{BIJC}$.
\end{theorem}

\subsection{Discussion: Convergence Speed/The Limits}

For any given graph signal $\mathrm{G}_{\sigma(t)}$, the value of $\mathbf{y}^\flat(\mathbf{x}(0))$ in Theorems \ref{thm3} and \ref{thm4}  depends only on the initial value $\mathbf{x}(0)$. We manage to provide a characterization to  $\mathbf{y}^\flat(\mathbf{x}(0))$ for balanced switching graphs or fixed graphs. Denote $\mathpzc{P}_{\mathcal{A}}$ as the projection operator over $\mathcal{A}$.

\begin{theorem}\label{prop1} The following statements hold for both the ``consensus + projection" and the ``projection consensus" flows.
\begin{itemize}
\item[(i)] Suppose $\mathrm{G}_{\sigma(t)}$ is balanced, i.e., $\sum_{j\in \mathrm{N}_i(t)} a_{ij}(t)=\sum_{i\in\mathrm{N}_j(t)}a_{ji}(t)$ for all $t\geq 0$ and  for all $i,j\in\mathrm{V}$. Suppose in addition  that $\mathrm{G}_{\sigma(t)}$ is either  $\delta$-$\mathsf{UJSC}$ or $\delta$-$\mathsf{BIJC}$. Then
$$
\lim_{t\to\infty} \mathbf{x}_i(t)=\sum_{i=1}^N \mathpzc{P}_{\mathcal{A}}\big(\mathbf{x}_i(0)\big)/N, \ \forall i\in\mathrm{V}.
$$
\item[(ii)] Suppose $\mathrm{G}_{\sigma(t)}\equiv\mathrm{G}^\S$ for some fixed, strongly connected, digraph $\mathrm{G}^\S$ and for any $i,j\in\mathrm{V}$, $a_{ij}(t)\equiv a_{ij}^\S$ for some constant $a_{ij}^\S$.  Let $\mathbf{w}:=(w_1 \dots w_N)\T$ with $\sum_{i=1}^N w_i=1$ be the left eigenvector  corresponding to the simple eigenvalue zero of the Laplacian\footnote{The Laplacian matrix $\mathbf{L}^\S$ associated with the graph $\mathrm{G}^\S$ under the given arc weights is defined as $\mathbf{L}^\S=\mathbf{D}^\S-\mathbf{A}^\S$ where $
\mathbf{A}^\S=[\mathbb{I}_{(j,i)\in\mathrm{E}^\S}a_{ij}^\S]
$ and $
\mathbf{D}^\S={\rm diag}\big(\sum_{j=1}^N \mathbb{I}_{(j,1)\in\mathrm{E}^\S}a_{1j}^\S, \dots, \sum_{j=1}^N \mathbb{I}_{(j,N)\in\mathrm{E}^\S}a_{Nj}^\S\big).
$ In fact, we have $w_i>0$ for all $i\in \mathrm{V}$ if $\mathrm{G}^\S$ is strongly connected \cite{magnusbook}.} $\mathbf{L}^\S$ of the digraph $\mathrm{G}^\S$.
Then we have
$$
\lim_{t\to\infty} \mathbf{x}_i(t)=\sum_{i=1}^N w_i\mathpzc{P}_{\mathcal{A}}\big(\mathbf{x}_i(0)\big),\  \forall i\in\mathrm{V}.
$$

\end{itemize}
\end{theorem}


 Due to the linear nature of the systems (\ref{1}) and (\ref{2}), it is straightforward to see that  the convergence stated in Theorems \ref{thm1}, \ref{thm2}, \ref{thm3} and \ref{thm4} is   exponential if $\mathrm{G}_{\sigma(t)}$ is periodic and  $\delta$-$\mathsf{UJSC}$. It is however difficult to provide tight approximations of the exact convergence rates because of the general  structure with switching interactions adopted by our network model. For time-invariant networks with constant edge weights, the two flows become linear time-invariant,  and then  the convergence rate is determined by the spectrum of the state transition matrix: the rate of convergence for the  ``consensus + projection" flow is determined together by both the graph Laplacian and the structure of the linear manifolds $\mathcal{A}_i$ and one in fact cannot dominate another;  the rate of convergence for the ``projection consensus" flow is upper bounded by the smallest nonzero eigenvalue of the graph Laplacian \cite{brian15}.

\subsection{Preliminaries and Auxiliary Lemmas}
Before presenting the detailed proofs for the stated results, we recall  some preliminary theories from  affine subspaces, convex analysis,  invariant sets, and  Dini derivatives, as well as a few auxiliary lemmas which will be useful  for the analysis.

\subsubsection{Affine Spaces and  Projection Operators}

An affine space is a set $\mathcal{X}$ that admits a free transitive action of a vector space $\mathcal{V}$. A set $\mathcal{K}\subset \mathbb{R}^d$ is said to be {\it convex} if $(1-\lambda)\mathbf{x}+\lambda
\mathbf{y}\in K$ whenever $\mathbf{x}\in \mathcal{K},\mathbf{y}\in \mathcal{K}$ and $0\leq\lambda \leq1$ \cite{rock}.
For any set $\mathcal{S}\subset \mathbb{R}^d$, the intersection of all convex sets
containing $\mathcal{S}$ is called the {\it convex hull} of $\mathcal{S}$, denoted by
${\rm co}(\mathcal{S})$. Let $\mathcal{K}$ be a closed convex subset in $\mathbb{R}^d$ and denote
$\|\mathbf{x}\|_\mathcal{K}\doteq\min_{\mathbf{y}\in \mathcal{K}} \| \mathbf{x}-\mathbf{y} \|$ as the distance between $\mathbf{x}\in \mathbb{R}^d$ and $\mathcal{K}$.  There is a unique element $\mathpzc{P}_{\mathcal{K}}(\mathbf{x})\in \mathcal{K}$ satisfying
$\|\mathbf{x}-\mathpzc{P}_{\mathcal{K}}(\mathbf{x})\|=\|\mathbf{x}\|_{\mathcal{K}}$ associated to any
$\mathbf{x}\in \mathbb{R}^d$ \cite{aubin}.  The map $\mathpzc{P}_{\mathcal{K}}$ is called the {\it projector} onto $\mathcal{K}$\footnote{The projections $\mathpzc{P}_{\mathcal{A}_i}$ and $\mathpzc{P}_{\mathcal{A}}$ introduced earlier are consistent with this definition.}. The following lemma holds  \cite{aubin}.
\begin{lemma}\label{lemconvex}
(i) $\langle\mathpzc{P}_{\mathcal{K}}(\mathbf{x})-\mathbf{x},\mathpzc{P}_{\mathcal{K}}(\mathbf{x})-\mathbf{y}\rangle\leq 0,\quad \forall  \mathbf{x}\in \mathbb{R}^d, \mathbf{y}\in
\mathcal{K}$.

(ii) $\|\mathpzc{P}_{\mathcal{K}}(\mathbf{x})-\mathpzc{P}_{\mathcal{K}}(\mathbf{y})\|\leq\|\mathbf{x}-\mathbf{y}\|, \ \forall \mathbf{x},\mathbf{y}\in \mathbb{R}^d$.

(iii)  The function $\|\mathbf{x}\|_\mathcal{K}^2$ is continuously differentiable at every $\mathbf{x}\in\mathbb{R}^d$ with $\nabla\|\mathbf{x}\|_\mathcal{K}^2=2\big(\mathbf{x}-\mathpzc{P}_{\mathcal{K}}(\mathbf{x})\big)$.
\end{lemma}

\subsubsection{Invariant Sets and Dini Derivatives}
Consider the following
autonomous system
\begin{equation}
\label{i1} \dot{\mathbf{x}}=f(\mathbf{x}),
\end{equation}
where $f:\mathbb{R}^d\rightarrow \mathbb{R}^d$ is a  continuous function. Let $\mathbf{x}(t)$ be a solution of
(\ref{i1}) with initial condition $\mathbf{x}(t_0)=\mathbf{x}^0$. Then $\Omega_0\subset \mathbb{R}^d$ is called a {\it positively invariant
set} of (\ref{i1}) if, for any $t_0\in\mathbb{R}$ and any $\mathbf{x}^0\in\Omega_0$,
we have $\mathbf{x}(t)\in\Omega_0$, $t\geq t_0$, along  every solution $\mathbf{x}(t)$ of (\ref{i1}).

The  {\it upper Dini
derivative} of a continuous function $h: (a,b)\to \mathbb{R}$ ($-\infty\leq a<b\leq \infty$) at $t$ is defined as
$$
D^+h(t)=\limsup_{s\to 0^+} \frac{h(t+s)-h(t)}{s}.
$$
When $h$ is continuous on $(a,b)$, $h$ is
non-increasing on $(a,b)$ if and only if $ D^+h(t)\leq 0$ for any
$t\in (a,b)$. Hence if $ D^+h(t)\leq 0$ for all $t\geq 0$ and $h(t)$ is lower bounded then the limit of $h(t)$ exists as a finite number when $t$ tends to infinity.

The next
result is convenient for the calculation of the Dini derivative \cite{dan}.

\begin{lemma}
\label{lemdini}  Let $V_i(t,x): \mathbb{R}\times \mathbb{R}^d \to \mathbb{R}\;(i=1,\dots,n)$ be
 a continuously differentiable  function with respect to both $t$ and $x$  and $V(t,x)=\max_{i=1,\dots,n}V_i(t,x)$. Let $x(t)$ be an absolutely continuous function. If $
\mathcal{I}(t)=\{i\in \{1,2,\dots,n\}\,:\,V(t,x(t))=V_i(t,x(t))\}$
is the set of indices where the maximum is reached at $t$, then
$
D^+V(t,x(t))=\max_{i\in\mathcal{ I}(t)}\dot{V}_i(t,x(t)).
$
\end{lemma}

\subsubsection{Key Lemmas}

The following lemmas, Lemmas \ref{lem:affineprojection}, \ref{lem:subspace-affine-projection}, \ref{lem:doubleprojection} can be easily proved using the properties of affine spaces, which turn out to be useful throughout our analysis. We therefore collect them below and the details of their proofs are omitted.
\begin{lemma}\label{lem:affineprojection}
Let $\mathcal{K}:=\{\mathbf{y}\in\mathbb{R}^m:\ \mathbf{a}\T \mathbf{y}=b\}$ be an affine subspace, where $\mathbf{a} \in \mathbb{R}^m$ with $\|\mathbf{a}\|=1$, and $b\in\mathbb{R}$. Denote $\mathpzc{P}_{\mathcal{K}}(\cdot): \mathbb{R}^m \rightarrow \mathcal{K}$ as the projection onto $\mathcal{K}$. Then $\mathpzc{P}_{\mathcal{K}}(\mathbf{y})=(I-\mathbf{a}\mathbf{a}\T)\mathbf{y}+b\mathbf{a}$ for all $\mathbf{y}\in\mathbb{R}^m$. Consequently there holds $\mathpzc{P}_{\mathcal{K}} (\mathbf{y} -\mathbf{u})=\mathpzc{P}_{\mathcal{K}}(\mathbf{y})-\mathpzc{P}_{\mathcal{K}}(\mathbf{u})+\mathpzc{P}_{\mathcal{K}}(0)$ for all $\mathbf{y},\mathbf{u}\in\mathbb{R}^m$.
\end{lemma}
\begin{lemma}\label{lem:subspace-affine-projection}
Let $\mathcal{K}$ be an affine subspace in $\mathbb{R}^m$. Take an arbitrary point $\mathbf{k}\in\mathcal{K}$ and define $\mathcal{Y}_{\mathbf{k}}:=\{\mathbf{y}-\mathbf{k}:\mathbf{y}\in\mathcal{K}\}$, which is a subspace in $\mathbb{R}^m$.  Denote $\mathpzc{P}_{\mathcal{K}}(\cdot): \mathbb{R}^m \rightarrow \mathcal{K}$  and $\mathpzc{P}_{\mathcal{Y}_{\mathbf{k}}}(\cdot): \mathbb{R}^m \rightarrow \mathcal{Y}_{\mathbf{k}}$ as the projectors onto $\mathcal{K}$ and $\mathcal{Y}_{\mathbf{k}}$, respectively. Then $\mathpzc{P}_{\mathcal{Y}_{\mathbf{k}}} (\mathbf{y} -\mathbf{u})=\mathpzc{P}_{\mathcal{K}}(\mathbf{y})-\mathpzc{P}_{\mathcal{K}}(\mathbf{u})$ for all $\mathbf{y},\mathbf{u}\in\mathbb{R}^m$.
\end{lemma}
\begin{lemma}\label{lem:doubleprojection}
Let $\mathcal{K}_1$ and $\mathcal{K}_2$ be two affine subspaces in $\mathbb{R}^m$ with $\mathcal{K}_1 \subseteq \mathcal{K}_2$. Denote $\mathpzc{P}_{\mathcal{K}_1}(\cdot): \mathbb{R}^m \rightarrow \mathcal{K}_1$ and  $\mathpzc{P}_{\mathcal{K}_2}(\cdot): \mathbb{R}^m \rightarrow \mathcal{K}_2$ as the projections onto $\mathcal{K}_1$ and $\mathcal{K}_2$, respectively. Then $
\mathpzc{P}_{\mathcal{K}_1}(\mathbf{y}) =\mathpzc{P}_{\mathcal{K}_1}\Big(\mathpzc{P}_{\mathcal{K}_2}(\mathbf{y})\Big), \ \forall \mathbf{y}\in\mathbb{R}^m.
$
\end{lemma}
\begin{lemma}\label{lem:invariant}
Suppose either (I) or (II) holds. Take $\mathbf{y}^\sharp$ as an arbitrary solution of (\ref{LinearEquation}) and let $r>0$ be arbitrary.  Define
$
\mathcal{M}^\sharp(r):=\Big\{\mathbf{w}\in\mathbb{R}^m:\ \big\|\mathbf{w}-\mathbf{y}^\sharp\big\|\leq r\Big\}.
$
Then \begin{itemize}
\item[(i)] $\big(\mathcal{M}^\sharp(r)\big)^{N}=\mathcal{M}^\sharp(r)\times \cdots \times \mathcal{M}^\sharp(r)$ is a positively invariant set for the ``consensus + projection" flow (\ref{1});
    \item[(ii)]  $\big(\mathcal{M}^\sharp(r)\big)^{N}\mcap\big(\mathcal{A}_1\times \cdots \times \mathcal{A}_N\big)$ is a positively invariant set for the ``projection consensus" flow (\ref{2}).
\end{itemize}
\end{lemma}
{\it Proof.} Let $\mathbf{x}(t)=(\mathbf{x}_1\T(t)\dots x_N\T(t))\T$ be a trajectory along the flows (\ref{1}) or (\ref{2}).  Define $
f^\sharp(t):=\max_{i\in\mathrm{V}}\frac{1}{2}\big\|\mathbf{x}_i(t)-\mathbf{y}^\sharp\big\|^2.
$
Denote $\mathcal{I}(t):=\big\{i: f^\sharp(t)=\big\|\mathbf{x}_i(t)-\mathbf{y}^\sharp\big\|^2\big\}$.

\noindent (i) From Lemma \ref{lemdini} we obtain that along the flow (\ref{1})
\begin{align}\label{3}
&D^+f^\sharp(t)=\max_{i\in\mathcal{I}(t)} \Big\langle \mathbf{x}_i(t)-\mathbf{y}^\sharp,  K\Big( \sum_{j\in \mathrm{N}_i(t)}a_{ij}(t)\big(\mathbf{x}_j(t)-\mathbf{x}_i(t)\big) \Big) + \mathpzc{P}_{\mathcal{A}_i}(\mathbf{x}_i(t))-\mathbf{x}_i(t)\Big\rangle\nonumber\\
&\stackrel{a)}{=}\max_{i\in\mathcal{I}(t)}\bigg[ K \sum_{j\in \mathrm{N}_i(t)}a_{ij}(t) \Big\langle \mathbf{x}_i(t)-\mathbf{y}^\sharp, \big(\mathbf{x}_j(t)-\mathbf{y}^\sharp\big)-\big(\mathbf{x}_i(t)-\mathbf{y}^\sharp\big) \Big\rangle-\big|\mathbf{h}_i^T( \mathbf{x}_i(t)-\mathbf{y}^\sharp)\big|^2 \bigg]\nonumber\\
&\stackrel{b)}{\leq}  \max_{i\in\mathcal{I}(t)}\bigg[ K \sum_{j\in \mathrm{N}_i(t)}\frac{a_{ij}(t)}{2}  \Big(\big\|\mathbf{x}_j(t)-\mathbf{y}^\sharp\big\|^2 - \big\|\mathbf{x}_i(t)-\mathbf{y}^\sharp\big\|^2 \Big)-\big|\mathbf{h}_i^T( \mathbf{x}_i(t)-\mathbf{y}^\sharp)\big|^2 \bigg]\nonumber\\
&\stackrel{c)}{\leq}   \max_{i\in\mathcal{I}(t)} \bigg[-\big|\mathbf{h}_i^T( \mathbf{x}_i(t)-\mathbf{y}^\sharp)\big|^2 \bigg]\nonumber\\
&\leq 0,
\end{align}
where $a)$ follows from the fact that
$
\big\langle \mathbf{x}_i(t)-\mathbf{y}^\sharp, \mathpzc{P}_{\mathcal{A}_i}(\mathbf{x}_i(t))-\mathbf{x}_i(t)\big\rangle
=-\big|\mathbf{h}_i^T( \mathbf{x}_i(t)-\mathbf{y}^\sharp)\big|^2
$
in view of Lemma \ref{lem:affineprojection} and the fact that $\mathbf{y}^\sharp$ is a solution of (\ref{LinearEquation}); $b)$ makes use of the elementary inequality
\begin{align}\label{4}
\langle \alpha, \beta\rangle\leq \big(\|\alpha\|^2+ \|\beta\|^2\big)/2,\ \forall \alpha,\beta\in \mathbb{R}^m;
\end{align}
$c)$ follows from the definition of $f^\sharp(t)$ and $\mathcal{I}(t)$.

We therefore know from (\ref{3}) that $f^\sharp(t)$ is always a non-increasing functions. This implies that,  if $\mathbf{x}(t_0)\in\big(\mathcal{M}^\sharp(r)\big)^{N}$, i.e., $\big\|\mathbf{x}_i(t_0)-\mathbf{y}^\sharp\big\|\leq r$  for all $i$, then $\big\|\mathbf{x}_i(t)-\mathbf{y}^\sharp\big\|\leq r$  for all $i$ and all $t\geq t_0$. Equivalently,  $\big(\mathcal{M}^\sharp(r)\big)^{N}$ is a positively invariant set for the flow (\ref{1}).

\noindent(ii) Let $\mathbf{x}(t_0)\in\mathcal{A}_1\times \cdots \times \mathcal{A}_N$. The structure of the flow (\ref{2}) immediately tells that $\mathbf{x}(t)\in\mathcal{A}_1\times \cdots \times \mathcal{A}_N$ for all $t\geq t_0$.  Furthermore, again by Lemma \ref{lemdini} we obtain
\begin{align}\label{8}
&D^+f^\sharp(t)=\max_{i\in\mathcal{I}(t)} \Big\langle \mathbf{x}_i(t)-\mathbf{y}^\sharp, \sum_{j\in \mathrm{N}_i(t)}a_{ij}(t)\big(\mathpzc{P}_{\mathcal{A}_i}(\mathbf{x}_j)-\mathpzc{P}_{\mathcal{A}_i}(\mathbf{x}_i)\big)\Big\rangle\nonumber\\
&\stackrel{a)}{=}\max_{i\in\mathcal{I}(t)}\  \sum_{j\in \mathrm{N}_i(t)}a_{ij}(t) \Big\langle \mathbf{x}_i(t)-\mathbf{y}^\sharp,  \big(I-\mathbf{h}_i\mathbf{h}_i^T\big) \big(\mathbf{x}_j(t)-\mathbf{x}_i(t)\big)\Big\rangle  \nonumber\\
&\stackrel{b)}{=}\max_{i\in\mathcal{I}(t)}  \sum_{j\in \mathrm{N}_i(t)}a_{ij}(t) \Big( \mathbf{x}_i(t)-\mathbf{y}^\sharp\Big)\T \Big(I-\mathbf{h}_i\mathbf{h}_i^T\Big)^2  \cdot\Big(\big(\mathbf{x}_j(t)-\mathbf{y}^\sharp\big)-\big(\mathbf{x}_i(t)-\mathbf{y}^\sharp\big)\Big)\nonumber\\
&\stackrel{c)}{\leq}  \max_{i\in\mathcal{I}(t)}\bigg[  \sum_{j\in \mathrm{N}_i(t)}\frac{a_{ij}(t)}{2}  \Big(\Big\|\Big(I-\mathbf{h}_i\mathbf{h}_i^T\Big)\Big(\mathbf{x}_j(t)-\mathbf{y}^\sharp\Big)\Big\|^2 - \Big\|\Big(I-\mathbf{h}_i\mathbf{h}_i^T\Big)\Big(\mathbf{x}_i(t)-\mathbf{y}^\sharp\Big)\Big\|^2  \Big)\bigg]\nonumber\\
&\stackrel{d)}{\leq}  \max_{i\in\mathcal{I}(t)}\bigg[  \sum_{j\in \mathrm{N}_i(t)}\frac{a_{ij}(t)}{2}  \Big(\big\|\mathbf{x}_j(t)-\mathbf{y}^\sharp\big\|^2 - \big\|\mathbf{x}_i(t)-\mathbf{y}^\sharp\big\|^2  \Big)\bigg]\nonumber\\
&\leq 0,
\end{align}
where $a)$ follows from Lemma \ref{lem:affineprojection}, $b)$ holds due to the fact that $I-\mathbf{h}_i\mathbf{h}_i^T$ is a projection  matrix, $c)$ is again based on the inequality (\ref{4}), and $d)$ holds because $x_i(t)\in \mathcal{A}_i$ (so that $\big(I-\mathbf{h}_i\mathbf{h}_i^T\big)\big(\mathbf{x}_i(t)-\mathbf{y}^\sharp\big)=\mathbf{x}_i(t)-\mathbf{y}^\sharp$) and   that  $I-\mathbf{h}_i\mathbf{h}_i^T$ is a projection  matrix (so that $\big\|\big(I-\mathbf{h}_i\mathbf{h}_i^T\big)\big(\mathbf{x}_j(t)-\mathbf{y}^\sharp\big)\big\|\leq \big\|\mathbf{x}_j(t)-\mathbf{y}^\sharp\big\|$).

We can therefore readily conclude that $\big(\mathcal{M}^\sharp(r)\big)^{N}\mcap\big(\mathcal{A}_1\times \cdots \times \mathcal{A}_N\big)$ is a positively invariant set for the ``projection consensus" flow (\ref{2}). \hfill$\square$

\subsection{Proofs of Statements}
We  now have the tools in place to present the proofs of the stated results.
\subsubsection{Proof of Theorems \ref{thm1} and \ref{thm2}}
When (I) holds, $
\mathcal{A}:= \mcap_{i=1}^N\mathcal{A}_i=\{\mathbf{y}_\ast\}
$ is a singleton, which is obviously a bounded set. The ``consensus + projection" flow (\ref{1}) is a special case of the optimal consensus flow proposed in \cite{shitac}. Theorem \ref{thm1} readily follows  from Theorems 3.1 and Theorems 3.2 in \cite{shitac} by adapting the treatments to the Weights Assumption adopted  in the current paper using the techniques established in \cite{shisiam}. We therefore omit the details.

Being a special case of  Theorem \ref{thm4},  Theorem \ref{thm2} holds true as a direct consequence of Theorem \ref{thm4}, whose proof will be presented below.

\subsubsection{Proof of Theorem \ref{thm3}}
Recall that $\mathbf{y}^\sharp$  is  an arbitrary solution of (\ref{LinearEquation}). With Lemma \ref{lem:invariant}, for any initial value $\mathbf{x}(0)$,   the set
$$
\big(\mathcal{M}^\sharp(\max_{i\in\mathrm{V}}\|\mathbf{x}_i(0)\|)\big)^{N},
$$
which  is  obviously bounded, is a positively invariant set for the ``consensus + projection" flow (\ref{1}). Define $$
\mathcal{A}_i^\sharp:=\mathcal{A}_i \mcap\mathcal{M}^\sharp(\max_{i\in\mathrm{V}}\|\mathbf{x}_i(0)\|),\ i=1,\dots,N
$$
and $\mathcal{A}^\sharp:=\mathcal{A} \mcap\mathcal{M}^\sharp(\max_{i\in\mathrm{V}}\|\mathbf{x}_i(0)\|)$.  As a result,  recalling  Theorem 3.1 and Theorem 3.2 from \cite{shitac}\footnote{Again, the arguments in \cite{shitac} were based on switching graph signals with dwell time and absolutely bounded weight functions. We can however borrow the treatments in \cite{shisiam} to the generalized graph and arc weight model considered here under which we can rebuild the results in \cite{shitac}. More details will be provided in the proof of Theorem \ref{thm4}.   }, if $\mathrm{G}_{\sigma(t)}$ is either  $\delta$-$\mathsf{UJSC}$ or $\delta$-$\mathsf{BIJC}$, there hold

(i) $\lim_{t\to \infty} \|\mathbf{x}_i(t)\|_{\mathcal{A}^\sharp}=0$ for all $i\in\mathrm{V}$;

(ii) $\lim_{t\to \infty} \|\mathbf{x}_i(t)-\mathbf{x}_j(t)\|=0$  for all $i$ and $j$.

 We still need to show that the limits of the $\mathbf{x}_i(t)$ indeed exist. Introduce
\begin{align*}
\mathcal{S}_i:=\Big\{\mathbf{y}:\ \mathbf{h}_i\T \mathbf{y}=0 \Big\},\ i\in\mathrm{V}
\end{align*}
and $\mathcal{S}:=\mcap_{i=1}^N \mathcal{S}_i$.

 With Lemma \ref{lem:doubleprojection},  we see that
 \begin{align}
 &\mathpzc{P}_{\mathcal{S}} \Big( \mathpzc{P}_{\mathcal{S}_i}(\mathbf{x}_i-\mathbf{y}^\sharp)\Big)-\mathpzc{P}_{\mathcal{S}} \big(\mathbf{x}_i-\mathbf{y}^\sharp\big)=\mathpzc{P}_{\mathcal{S}} \big(\mathbf{x}_i-\mathbf{y}^\sharp\big)-\mathpzc{P}_{\mathcal{S}} \big(\mathbf{x}_i-\mathbf{y}^\sharp\big)=0.
 \end{align}
 As a result, taking $\mathpzc{P}_{\mathcal{S}}(\cdot)$ from both the left and right sides of (\ref{1})\footnote{Note that, $\mathpzc{P}_{\mathcal{S}}(\cdot)$ is a projector onto a subspace.}, we obtain
\begin{align}\label{20}
&\frac{d}{dt} \mathpzc{P}_{\mathcal{S}}(\mathbf{x}_i(t)-\mathbf{y}^\sharp)\nonumber\\
&= K\Big( \sum_{j\in \mathrm{N}_i(t)}a_{ij}(t)\big(\mathpzc{P}_{\mathcal{S}}(\mathbf{x}_j(t)-\mathbf{y}^\sharp)-\mathpzc{P}_{\mathcal{S}}(\mathbf{x}_i(t)-\mathbf{y}^\sharp)\big) \Big) + \mathpzc{P}_{\mathcal{S}} \Big( \mathpzc{P}_{\mathcal{S}_i}(\mathbf{x}_i-\mathbf{y}^\sharp)\Big)-\mathpzc{P}_{\mathcal{S}} \big(\mathbf{x}_i-\mathbf{y}^\sharp\big)\nonumber\\
&= K\Big( \sum_{j\in \mathrm{N}_i(t)}a_{ij}(t)\big(\mathpzc{P}_{\mathcal{S}}(\mathbf{x}_j(t)-\mathbf{y}^\sharp)-\mathpzc{P}_{\mathcal{S}}(\mathbf{x}_i(t)-\mathbf{y}^\sharp)\big) \Big).
\end{align}
This is to say, if $\mathrm{G}_{\sigma(t)}$ is either  $\delta$-$\mathsf{UJSC}$ or $\delta$-$\mathsf{BIJC}$, we can invoke Theorem 4.1 (for $\delta$-$\mathsf{UJSC}$ graphs) and Theorem 5.2 (for $\delta$-$\mathsf{BIJC}$ graphs) in \cite{shisiam} to conclude:  for any initial value $\mathbf{x}(0)$, there exists $\mathbf{p}^0 (\mathbf{x}(0))\in\mathcal{S}$ such that
\begin{align}\label{5}
\lim_{t\rightarrow \infty}\mathpzc{P}_{\mathcal{S}}(\mathbf{x}_i(t)-\mathbf{y}^\sharp)=\mathbf{p}^0,\ \ i=1,\dots,N.
\end{align}
While on the other hand $\lim_{t\rightarrow \infty}\|\mathbf{x}_i(t)\|_{\mathcal{A}}\leq \lim_{t\to \infty} \|\mathbf{x}_i(t)\|_{\mathcal{A}^\sharp}=0$ leads to
\begin{align}\label{6}
0&=\lim_{t\rightarrow \infty}
\Big\| \mathbf{x}_i(t)-\mathpzc{P}_{\mathcal{A}}(\mathbf{x}_i(t))\Big\|\nonumber\\
&=\lim_{t\rightarrow \infty}
\Big\| \mathbf{x}_i(t)-\mathbf{y}^\sharp-\Big(\mathpzc{P}_{\mathcal{A}}(\mathbf{x}_i(t))-\mathbf{y}^\sharp\Big)\Big\|\nonumber\\
&=\lim_{t\rightarrow \infty}
\Big\| \mathbf{x}_i(t)-\mathbf{y}^\sharp-\mathpzc{P}_{\mathcal{S}}(\mathbf{x}_i(t)-\mathbf{y}^\sharp\big)\Big\|,
\end{align}
where the last equality follows from Lemma \ref{lem:subspace-affine-projection}. We conclude from (\ref{5}) and (\ref{6}) that
\begin{align}\label{21}
\lim_{t\to\infty} \mathbf{x}_i(t)=\mathbf{p}^0+\mathbf{y}^\sharp:=\mathbf{y}^\flat(\mathbf{x}(0)),\ i\in\mathcal{V}.
\end{align}
We have now completed the proof of Theorem \ref{thm3}.
\subsubsection{Proof of Theorem \ref{thm4}}
We continue to use the notation   $$
\mathcal{A}_i^\sharp:=\mathcal{A}_i \mcap\mathcal{M}^\sharp(\max_{i\in\mathrm{V}}\|\mathbf{x}_i(0)\|),\ i=1,\dots,N
$$
and $\mathcal{A}^\sharp:=\mathcal{A} \mcap\mathcal{M}^\sharp(\max_{i\in\mathrm{V}}\|\mathbf{x}_i(0)\|)$ introduced in the proof of Theorem \ref{thm3}. In view of Lemma \ref{lem:invariant}, for any initial value $\mathbf{x}(0)$,
 $$
\mathcal{A}_1^\sharp\times \cdots \times \mathcal{A}_N^\sharp
$$ is a positively invariant set for the ``projection consensus" flow (\ref{2}). Define
\begin{align}
h^\sharp(t):= \max_{i\in\mathrm{V}}\frac{1}{2}\big\|\mathbf{x}_i(t)\big\|_{\mathcal{A}^\sharp}^2.
\end{align}

Let $\mathcal{I}_0(t)=\big\{i: h^\sharp(t)= \big\|\mathbf{x}_i(t)\big\|_{\mathcal{A}^\sharp}^2\big\}$.  We obtain from Lemma \ref{lemconvex}.(iii) that
\begin{align}\label{7}
D^+h^\sharp(t)&=\max_{i\in\mathcal{I}_0(t)} \Big\langle \mathbf{x}_i(t)-\mathpzc{P}_{\mathcal{A}^\sharp}(\mathbf{x}_i(t)), \sum_{j\in \mathrm{N}_i(t)}a_{ij}(t)\big(\mathpzc{P}_{\mathcal{A}_i}(\mathbf{x}_j)-\mathpzc{P}_{\mathcal{A}_i}(\mathbf{x}_i)\big)\Big\rangle.
\end{align}
In (\ref{7}), the argument used to establish (\ref{8}) can be carried through when $\mathbf{y}^\sharp$  is replaced by  $\mathpzc{P}_{\mathcal{A}^\sharp}(\mathbf{x}_i(t))$. Accordingly, we obtain\footnote{Note that the inequalities in (\ref{8}) hold without relying   on the fact that $\mathbf{y}^\sharp$ is a constant.}, $D^+h^\sharp(t)\leq 0$. This immediately implies that   there is a  constant $h_\S\geq 0$ such that $
\lim_{t\to \infty}h^\sharp(t)=h_\S^2/2$. As a result, for any $\epsilon>0$, there exists $T_{\epsilon}>0$ such that
\begin{align}\label{10}
\big\|\mathbf{x}_i(t)\big\|_{\mathcal{A}^\sharp}\leq h_\S+\epsilon,\ \forall t\geq  T_{\epsilon}.
\end{align}

Now that $h_\S$ is a nonnegative constant satisfying  $
\lim_{t\to \infty}h^\sharp(t)=h_\S^2/2$, we can in fact show that $h_\S=0$ with suitable graph conditions, as summarized in the following  two technical lemmas. Details of their proofs can be found  in the Appendix.

\medskip

\begin{lemma}\label{lem:directed}
Let  $\mathrm{G}_{\sigma(t)}$ be $\delta$-$\mathsf{UJSC}$. Then $h_\S=0$. In fact, $h_\S=0$ due to the following two contradictive conclusions:

  (i) If $h_\S>0$, then  there holds   $\lim_{t\to\infty}\|\mathbf{x}_i(t)\|_{\mathcal{A}^\sharp}=h_\S$ for all $i\in \mathrm{V}$  along the ``projection consensus" flow (\ref{2}).

  (ii) If    $\lim_{t\to\infty}\|\mathbf{x}_i(t)\|_{\mathcal{A}^\sharp}=h_\S$  for all $i\in \mathrm{V}$  along the ``projection consensus" flow (\ref{2}), then $h_\S=0$.
\end{lemma}

\medskip

\begin{lemma}\label{lemma-bidirectional}
Suppose $\mathrm{G}_{\sigma(t)}$ is $\delta$-$\mathsf{BIJC}$. Then $h_\S=0$ along the ``projection consensus" flow (\ref{2}).
\end{lemma}

\medskip

%

Recall  $\mathbf{y}^\sharp \in \mathcal{A}$. Applying $\mathpzc{P}_{\mathcal{S}}(\cdot)$ to both the left and right sides of (\ref{2}), we have
\begin{align}\label{25}
&\frac{d}{dt} \mathpzc{P}_{\mathcal{S}}({\mathbf{x}}_i-\mathbf{y}^\sharp)=  \sum_{j\in \mathrm{N}_i(t)}a_{ij}(t)\big((\mathpzc{P}_{\mathcal{S}}({\mathbf{x}}_j-\mathbf{y}^\sharp)-\mathpzc{P}_{\mathcal{S}}({\mathbf{x}}_i-\mathbf{y}^\sharp)\big)
\end{align}
where we have used Lemma \ref{lem:subspace-affine-projection} to derive
\begin{align}
\mathpzc{P}_{\mathcal{S}}\big(\mathpzc{P}_{\mathcal{A}_i}({\mathbf{x}}_j)-\mathbf{y}^\sharp\big)&= \mathpzc{P}_{\mathcal{A}}\big(\mathpzc{P}_{\mathcal{A}_i}({\mathbf{x}}_j)\big)-\mathbf{y}^\sharp= \mathpzc{P}_{\mathcal{A}}({\mathbf{x}}_j)-\mathbf{y}^\sharp=\mathpzc{P}_{\mathcal{S}}({\mathbf{x}}_j-\mathbf{y}^\sharp).
\end{align} We can again make use of the argument  in the proof of Theorem \ref{thm3} and conclude that the limits of the ${\mathbf{x}}_i(t)$ exist and they are obviously the same.

We have now completed the proof of Theorem \ref{thm4}.

\subsubsection{Proof of Theorem \ref{prop1}}

We provide detailed proof for the ``projection+consensus" flow.  The analysis for the projection consensus flow can be similarly established in view of  (\ref{25}).

\noindent (i) With (\ref{20}), we have \begin{align}\label{22}
&\frac{d}{dt} \mathpzc{P}_{\mathcal{S}}(\mathbf{x}_i(t)-\mathbf{y}^\sharp)= K\Big( \sum_{j\in \mathrm{N}_i(t)}a_{ij}(t)\big(\mathpzc{P}_{\mathcal{S}}(\mathbf{x}_j(t)-\mathbf{y}^\sharp)-\mathpzc{P}_{\mathcal{S}}(\mathbf{x}_i(t)-\mathbf{y}^\sharp)\big) \Big)
\end{align}
along the ``projection+consensus" flow. Note that (\ref{22}) is a standard consensus flow with arguments being $\mathpzc{P}_{\mathcal{S}}(\mathbf{x}_i(t)-\mathbf{y}^\sharp)$, $i=1,\dots,N$. As a result, there holds
\begin{align*}
\lim_{t\to\infty}\mathpzc{P}_{\mathcal{S}}(\mathbf{x}_i(t)-\mathbf{y}^\sharp)&= \sum_{i=1}^N \mathpzc{P}_{\mathcal{S}}(\mathbf{x}_i(0)-\mathbf{y}^\sharp)/N\nonumber\\
&=\sum_{i=1}^N \Big(\mathpzc{P}_{\mathcal{A}}(\mathbf{x}_i(0))-\mathbf{y}^\sharp\Big)/N
\end{align*}
if $\mathrm{G}_{\sigma(t)}$ is balanced for all $t$  \cite{saber04}. As a result, similar to  (\ref{21}), we have
\begin{align}
\lim_{t\to\infty} \mathbf{x}_i(t)&=\mathbf{p}^0+\mathbf{y}^\sharp\nonumber\\
&:=\sum_{i=1}^N \Big(\mathpzc{P}_{\mathcal{A}}(\mathbf{x}_i(0))-\mathbf{y}^\sharp\Big)/N+\mathbf{y}^\sharp\nonumber\\
&=\sum_{i=1}^N \mathpzc{P}_{\mathcal{A}}(\mathbf{x}_i(0))/N,\ i\in\mathrm{V}.
\end{align}

\noindent (ii) If $\mathrm{G}_{\sigma(t)}\equiv\mathrm{G}^\S$ for some fixed digraph $\mathrm{G}^\S$ and for any $i,j\in\mathrm{V}$, $a_{ij}(t)\equiv a_{ij}^\S$ for some constant $a_{ij}^\S$, then along (\ref{22}) we have   \cite{saber04}
\begin{align*}
\lim_{t\to\infty}\mathpzc{P}_{\mathcal{S}}(\mathbf{x}_i(t)-\mathbf{y}^\sharp)&= \sum_{i=1}^N w_i\mathpzc{P}_{\mathcal{S}}(\mathbf{x}_i(0)-\mathbf{y}^\sharp)=\sum_{i=1}^N w_i\mathpzc{P}_{\mathcal{A}}(\mathbf{x}_i(0))-\mathbf{y}^\sharp,
\end{align*}
where $\mathbf{w}:=(w_1 \dots w_N)\T$ is the left eigenvector corresponding to eigenvalue zero for the Laplacian matrix $\mathbf{L}^\S$. This immediately gives us
\begin{align}
\lim_{t\to\infty} \mathbf{x}_i(t)=\sum_{i=1}^N w_i\mathpzc{P}_{\mathcal{A}}(\mathbf{x}_i(0)),\ i\in\mathrm{V},
\end{align}
which completes the proof.
\section{Least Squares Solutions}\label{Sec:least-squares}

In this section, we turn to  Case  (III) and consider that (\ref{LinearEquation}) admits a unique least squares solution $\mathbf{y}^\star$. Evidently, neither of the two continuous-time distributed flows (\ref{1}) and (\ref{2}) in general can yield exact convergence to the least squares solution of (\ref{LinearEquation}) since, even for a fixed interaction graph, $\mathbf{y}^\star$ is not an equilibrium of the two network flows.

It is indeed possible to find the least squares solution using  double-integrator node dynamics \cite{elia,cotes14}.  However, the use of double integrator dynamics was  restricted to   networks with fixed and undirected (or balanced) communication graphs \cite{elia,cotes14}. In another direction,  one can also borrow the idea of  the use of    square-summable step-size sequences with infinite sums in discrete-time algorithms \cite{nedic10} and  build the following flow
\begin{align}\label{9}
\dot{\mathbf{x}}_i =K\Big( \sum_{j\in \mathrm{N}_i(t)}a_{ij}(t)\big(\mathbf{x}_j-\mathbf{x}_i\big) \Big)+ \frac{1}{t}\Big(\mathpzc{P}_{\mathcal{A}_i}(\mathbf{x}_i)-\mathbf{x}_i\Big),
\end{align}
for $i\in\mathrm{V}$.
The least squares case can then be solved under graph conditions of connectedness and balance \cite{nedic10}, but the convergence rate is at best $O(1/t)$. This means (\ref{9}) will be fragile against noises.

For the ``projection+consensus" flow, we can show that under fixed and connected bidirectional interaction graphs,  with a sufficiently large $K$, the node states will converge to  a ball around the least squares solution whose radius can be made arbitrarily small. This approximation is global in the sense that the required $K$ only depends on the accuracy between the node state limits and the $\mathbf{y}^\star$.
\begin{theorem}\label{thm5}
Let (III) hold with ${\rm rank}{(\mathbf{H})}=m$ and  denote the unique least squares solution of (\ref{LinearEquation}) as $\mathbf{y}^\star$.  Suppose $\mathrm{G}_{\sigma(t)}=\mathrm{G}^\S=(\mathrm{V}, \mathrm{E}^\S)$ for some bidirectional, connected graph $\mathrm{G}^\S$ and for any $i,j\in\mathrm{V}$, $a_{ij}(t)=a_{ji}(t)\equiv a_{ij}^\S$ for some constant $a_{ij}^\S$. Then   along  the flow (\ref{1}), for any $\epsilon>0$, there exists $K_\ast(\epsilon)>0$ such that $\mathbf{x}(\infty):=\lim_{t\to \infty}\mathbf{x}(t)$ exists and
 $$
 \big\|\mathbf{x}_i(\infty)-\mathbf{y}^\star\big\|\leq \epsilon,\  \forall i\in\mathrm{V}
$$
for any initial value $\mathbf{x}(0)$ if $K\geq K_\ast(\epsilon)$.
\end{theorem}

The intuition behind Theorem \ref{thm5} can be described briefly as follows (see its proof presented later for a full exposure of this intuition). With fixed and undirected network topology,  the ``consensus + projection" flow (\ref{1}) is a gradient flow  in the form of
$$
\dot{\mathbf{x}}_i=-\nabla_{\mathbf{x}_i} \mathpzc{D}_K(\mathbf{x})
$$
where
\begin{align}
\mathpzc{D}_K(\mathbf{x}):=\frac{1}{2}\sum_{i=1}^N \big \|\mathbf{x}_i\big\|_{\mathcal{A}_i}^2+ \frac{K}{2}\sum_{\{i,j\}\in\mathrm{E}^\S}a_{ij}^\S\big\|\mathbf{x}_j-\mathbf{x}_i\big\|^2.
\end{align}
There  holds
$
\sum_{i=1}^N \|\mathbf{x}_i\|_{\mathcal{A}_i}^2=\big|\mathbf{h}_i\T \mathbf{x}_i -z_i\big|^2
$
if each $\mathbf{h}_i$ is normalized with a unit length. Therefore, for large $K$, the trajectory of (\ref{1}) will asymptotically  tend to  be close to the solution of the following optimization problem:
\[
\begin{aligned}
\min_{\mathbf{y}_1,\dots, \mathbf{y}_{_N} \in \mathbb{R}^m}\qquad &  \sum_{i=1}^N \big|\mathbf{h}_i\T \mathbf{y}_i -z_i\big|^2 \\
{\rm s.t.} \qquad  &   \mathbf{y}_1=\dots=\mathbf{y}_{_N}.
\end{aligned}
\]
Any solution of the above problem  consists of  $N$ copies of the solution to
\[
\begin{aligned}
\min_{\mathbf{y} \in \mathbb{R}^m}\qquad &   \big\|\mathbf{H} \mathbf{y} -\mathbf{z}\big\|^2
\end{aligned}
\]
which is  the unique least squares  solution $\mathbf{y}^\star=(\mathbf{H}\T\mathbf{H})^{-1}\mathbf{H}\mathbf{z}$ if ${\rm rank}(\mathbf{H})=m$. Therefore, a large $K$ can drive the node states to somewhere near $\mathbf{y}^\star$ along the flow (\ref{1}) as stated in Theorem \ref{thm5}.

\begin{remark}[Least Squares Solutions without Normalizing $\mathbf{h}_i$]
From the above argument, for the case when $\mathbf{h}_i$ is not normalized, we can replace the ``consensus + projection" flow (\ref{1}) with
\begin{align}\label{r3}
 \dot{\mathbf{x}}_i
 &=K\Big( \sum_{j\in \mathrm{N}_i(t)}a_{ij}(t)\big(\mathbf{x}_j-\mathbf{x}_i\big) \Big)- \nabla_{\mathbf{x}_i}
\big|\mathbf{h}_i\T \mathbf{x}_i -z_i \big|^2/2 \nonumber\\
&=K\Big( \sum_{j\in \mathrm{N}_i(t)}a_{ij}(t)\big(\mathbf{x}_j-\mathbf{x}_i\big) \Big)- \mathbf{h}_i \big(\mathbf{h}\T_i\mathbf{x}_i -z_i \big)
\end{align}
 for all $i\in\mathrm{V}$. The statement  of Theorem \ref{thm5} will continue to hold for the flow (\ref{r3}).
\end{remark}

\begin{remark}[Tradeoff between Convergence Rate and Accuracy]
Under the assumptions of Theorem \ref{thm5}, the flow defines a  linear time-invariant system
\begin{align}
\dot{\mathbf{x}}=-\Big(K \mathbf{L}^\S\otimes \mathbf{I}_m +\mathbf{J}\Big)\mathbf{x}+{\mathbf{h}_z}
\end{align}
where $\mathbf{L}^\S$ is the Laplacian of the graph $\mathrm{G}^\S$,  $\mathbf{J}={\rm diag} (\mathbf{h}_1\mathbf{h}\T_1, \cdots, \mathbf{h}_N\mathbf{h}\T_N)$ is a block-diagonal matrix, and $\mathbf{h}_z=(z_1\mathbf{h}_1\T \cdots z_N \mathbf{h}_N\T)\T$. Therefore, the rate of convergence is given by
 $$
 \lambda_{\rm min}\Big(K \mathbf{L}^\S\otimes \mathbf{I}_m +\mathbf{J}\Big),
   $$
   which is influenced by both $\mathbf{L}^\S$ (the graph) and $\mathbf{J}$ (the linear equation). We also know that  $\lambda_{\rm min}\big(K \mathbf{L}^\S\otimes \mathbf{I}_m +\mathbf{J}\big)$ in general cannot grow to infinity by increasing $K$ due to the presence of the zero eigenvalue in $\mathbf{L}^\S$. One can however speed up the convergence by adding a weight $\gamma$ to the projection term in the flow (\ref{1}) and obtain
   \begin{align}\label{r9}
 \dot{\mathbf{x}}_i =K\Big( \sum_{j\in \mathrm{N}_i(t)}a_{ij}(t)\big(\mathbf{x}_j-\mathbf{x}_i\big) \Big)+ \gamma\Big( \mathpzc{P}_{\mathcal{A}_i}(\mathbf{x}_i)-\mathbf{x}_i\Big),\ \  i\in\mathrm{V}.
\end{align}
  Certainly the convergence rate (to a ball centered at $\mathbf{y}^\star$ with radius $\epsilon$) of (\ref{r9}) can be arbitrarily large  by selecting sufficiently large $K$ and $\gamma$. However, it is clear from the proof of Theorem \ref{thm5} that the required $K$ for a given accuracy $\epsilon$  will in turn depend on $\gamma$ and require a large $K$ for a large $\gamma$.

%

 \end{remark}

  We also manage to establish the following semi-global result for  switching but balanced graphs under two further assumptions.

\medskip

\noindent {\bf [A1]} The set $\mathcal{W}(\mathbf{y}):= \big\{\mathpzc{P}_{\mathcal{A}_{i_J}}\cdots \mathpzc{P}_{\mathcal{A}_{i_1}} (\mathbf{y}):\ i_1,\dots,i_J\in\mathrm{V}, J\geq 1\big\}$ is a bounded set.

\noindent {\bf [A2]} $\sum_{i=1}^N \mathpzc{P}_{\mathcal{A}_i}(0)=0$.

\begin{theorem}\label{thm6}
Let (III) hold with ${\rm rank}{(\mathbf{H})}=m$ and  denote the unique least squares solution of (\ref{LinearEquation}) as $\mathbf{y}^\star$. Assume [A1] and [A2].  Suppose $\mathrm{G}_{\sigma(t)}$ is balanced for all $t\in \mathbb{R}^+$ and   $\delta$-$\mathsf{UJSC}$ with respect to $T>0$. Then   for any $\epsilon>0$ and  any $\mathbf{x}(0)\in \mathcal{A}_1\times\cdots \times \mathcal{A}_N$, there exist two constants $K_\ast(\epsilon,\mathbf{x}(0))>0$ and $T_\ast(\epsilon,\mathbf{x}(0))$ such that
when $K\geq K_\ast$ and $T\leq T_\ast$, there holds $$
\limsup_{t\to \infty} \big\|\mathbf{x}_i(t)-\mathbf{y}^\star\big\|\leq \epsilon, \ \forall i\in\mathrm{V}
$$
along  the flow (\ref{1}) with the initial value $\mathbf{x}(0)$.
\end{theorem}

\begin{remark}
The two assumptions, [A1] and [A2] are indeed rather strong assumptions. In general, $[A1]$ holds if $\mcap_{i=1}^N \mathcal{A}_i$ is a nonempty bounded set \cite{B-B-SIAM}, which is exactly opposite to the least squares solution case under consideration. We conjecture that at least for $m=2$ case, $[A1]$ should hold  when $\mathbf{h}_1,\dots,\mathbf{h}_N$ are distinct vectors.  The assumption $[A2]$ requires a strong  symmetry in the affine spaces $\mathcal{A}_i$, which turns out to be essential for the result to stand.
\end{remark}

For the ``projection consensus" flow, we present the following result.

\begin{theorem}\label{thm7}
Let (III) hold with ${\rm rank}{(\mathbf{H})}=m$ and  denote the unique least squares solution of (\ref{LinearEquation}) as $\mathbf{y}^\star$. Suppose $\mathrm{G}_{\sigma(t)}=\mathrm{G}^\S$ is fixed, complete, and $a_{ij}(t) \equiv a^\S >0$ for all $i,j\in\mathrm{V}$.
Then    for any  initial value  $\mathbf{x}(0)\in \mathcal{A}_1\times\cdots \times \mathcal{A}_N$, there holds
\begin{align}
\lim_{t \to \infty} \frac{\sum_{i=1}^N \mathbf{x}_i(t)}{N} =\mathbf{y}^\star
\end{align}
along  the flow (\ref{2}).
\end{theorem}

 Based on simulation experiments,  the requirement that $\mathrm{G}_{\sigma(t)}=\mathrm{G}^\S$  does not have to be complete for the Theorem \ref{thm7} to be valid for certain choices of the $\mathcal{A}_i$. Moreover, because the flow (\ref{2}) is a gradient flow over the manifold $\mathcal{A}_1\times\cdots \times \mathcal{A}_N$, the $\mathbf{x}_i(t)$ will converge but perhaps to different limits at different nodes.
\subsection{Proof of Theorem \ref{thm5}}
 Suppose $\mathrm{G}_{\sigma(t)}=\mathrm{G}^\S$ for some bidirectional, connected graph $\mathrm{G}^\S$ and for any $i,j\in\mathrm{V}$, $a_{ij}(t)=a_{ji}(t)\equiv a_{ij}^\S$ for some constant $a_{ij}^\S$.  Fix $\epsilon>0$. Note that, we have
\begin{align}
K\sum\limits_{j \in
\mathrm{N}_i}a_{ij}^\S\big(\mathbf{x}_j-\mathbf{x}_i\big)+\mathpzc{P}_{\mathcal{A}_i}(\mathbf{x}_i)-\mathbf{x}_i=-\nabla_{\mathbf{x}_i} \mathpzc{D}_K(\mathbf{x}).
\end{align}

 Therefore, the flow (\ref{1}) is a gradient flow written as
$$
\dot{\mathbf{x}}= -\nabla \mathpzc{D}_K(\mathbf{x})
$$
where  $\mathpzc{D}_K(\mathbf{x})$ is a $C^{\infty}$ convex function.  Denote $
\mathcal{Z}_K:= \big\{\mathbf{x}: \nabla  \mathpzc{D}_K(\mathbf{x})=0 \big\}$. There must hold
\begin{align}\label{99}
\lim_{t\rightarrow \infty}\big\|\mathbf{x}(t)\big\|_{\mathcal{Z}_K}=0.
\end{align}

The following lemma holds.

\begin{lemma}
Suppose ${\rm rank}{(\mathbf{H})}=m$. Then  \begin{itemize}
\item[(i)] $\mathcal{Z}_K$ is a singleton, which implies that $\mathbf{x}(t)$ converges to a fixed point asymptotically;
\item[(ii)] $\mcup_{K>{\kappa_0}} \mathcal{Z}_K$ is a bounded set for all ${\kappa_0}>0$.
\end{itemize}
\end{lemma}
{\it Proof.} (i) Recall that $\mathbf{L}^\S$ is the Laplacian of the graph $\mathrm{G}^\S$,  $\mathbf{J}={\rm diag} (\mathbf{h}_1\mathbf{h}\T_1, \cdots, \mathbf{h}_N\mathbf{h}\T_N)$  is a block-diagonal matrix, and $\mathbf{h}_z=(z_1\mathbf{h}_1\T \cdots z_N \mathbf{h}_N\T)\T$. With Lemma \ref{lem:affineprojection},  the equation $\nabla_{\mathbf{x}} \mathpzc{D}_K(\mathbf{x})=0$ can be written as
\begin{align}\label{200}
K\sum_{j\in \mathrm{N}_i} a_{ij}^\S(\mathbf{x}_j -\mathbf{x}_i)-\mathbf{h}_i\mathbf{h}\T_i \mathbf{x}_i =z_i \mathbf{h}_i,\ i\in\mathrm{V},
\end{align}
or, in a compact form,
\begin{align}
\Big(K \mathbf{L}^\S\otimes \mathbf{I}_m +\mathbf{J}\Big) \mathbf{x}=-{\mathbf{h}_z}.
\end{align}

Consider
\begin{align*}
Q_K(\mathbf{x})&:=\mathbf{x}\T\Big(K \mathbf{L}^\S\otimes \mathbf{I}_m +\mathbf{J}\Big) \mathbf{x}\nonumber\\
&= \sum_{i=1}^N\big \|\mathbf{h}_i\T \mathbf{x}_i\big\|^2+ {K}\sum_{\{i,j\}\in\mathrm{E}^\S}a_{ij}^\S\big\|\mathbf{x}_j-\mathbf{x}_i\big\|^2.
\end{align*}
We immediately know from the second term of $Q$ that  $Q(\mathbf{x})=0$ only if $\mathbf{x}_1=\dots=\mathbf{x}_N$.  On the other hand,  obviously
$$
\sum_{i=1}^N\big |\mathbf{h}_i\T  \mathbf{w}\big|^2>0
$$
for any $\mathbf{w}\neq 0\in \mathbb{R}^m$ if ${\rm rank}{(\mathbf{H})}=m$. Therefore, $K \mathbf{L}^\S\otimes \mathbf{I}_m +\mathbf{J}$ is positive-definite, which yields
$$
\mathcal{Z}_K=\Big\{ - \Big( K \mathbf{L}^\S\otimes \mathbf{I}_m +\mathbf{J}\Big)^{-1} {\mathbf{h}_z}\Big\}.
$$

\noindent(ii) By the Courant-Fischer Theorem (see Theorem 4.2.11 in \cite{horn}), we have
\begin{align}
&\lambda_{\min}\Big( K \mathbf{L}^\S\otimes \mathbf{I}_m +\mathbf{J}\Big)=\min_{\|\mathbf{x}\|=1}Q_K(\mathbf{x})\nonumber\\
&=\min_{\|\mathbf{x}\|=1} \bigg[ \sum_{i=1}^N\big \|\mathbf{h}_i\T \mathbf{x}_i\big\|^2+ {K}\sum_{\{i,j\}\in\mathrm{E}}a_{ij}^\S\big\|\mathbf{x}_j-\mathbf{x}_i\big\|^2 \bigg].
\end{align}
This immediately implies
\begin{align}
\lambda_{\min}\Big( K \mathbf{L}^\S\otimes \mathbf{I}_m +\mathbf{J}\Big)\geq \lambda_{\min}\Big( {\kappa_0} \mathbf{L}^\S\otimes \mathbf{I}_m +\mathbf{J}\Big)>0,
\end{align}
for all $K\geq  {\kappa_0}$ and consequently,  $\mcup_{K>{\kappa_0}} \mathcal{Z}_K$ is obviously a bounded set. This proves the desired lemma. \hfill$\square$

Now we introduce  $\mathcal{Z}_\ast:=\mcup_{K\geq 1} \mathcal{Z}_K$. Let $\mathbf{w}=(\mathbf{w}_1\T \dots \mathbf{w}_N\T)\T$ with $\mathbf{w}_i\in\mathbb{R}^m$ and define
\begin{align}
{B}_0:= \sup_{\mathbf{w}\in \mathcal{Z}_\ast } \max_{i\in\mathrm{V}}\Big\|\mathpzc{P}_{\mathcal{A}_i}(\mathbf{w}_i)-\mathbf{w}_i\Big\|
 \end{align}
 and
 \begin{align}
{C}_0:= \sup_{\mathbf{w}\in \mathcal{Z}_\ast } \max_{i\in\mathrm{V}}\Big\|\mathbf{y}^\star-\mathbf{w}_i\Big\|
 \end{align}
 with $\mathbf{y}^\star$ being the unique least squares solution of (\ref{LinearEquation}). We see that $B_0$ and $C_0$ are finite numbers due to the boundedness of  $\mathcal{Z}_\ast$. The remainder of the proof contains two steps.

 \medskip

 \noindent{\bf Step 1}. Let $\mathbf{v}(K)= (\mathbf{v}_1\T(K) \dots \mathbf{v}_N(K)\T)=- \big( K \mathbf{L}^\S\otimes \mathbf{I}_m +\mathbf{J}\big)^{-1} {\mathbf{h}_z}\in\mathcal{Z}_K$. Then  $\mathbf{v}$  satisfies  (\ref{200}), or in equivalent form,
\begin{align}\label{32}
K \mathbf{L}^\S\otimes \mathbf{I}_m \mathbf{v} = \begin{pmatrix}
 \big(\mathbf{v}_1-\mathpzc{P}_{\mathcal{A}_1}(\mathbf{v}_1)\big)\T \\
 \big(\mathbf{v}_2-\mathpzc{P}_{\mathcal{A}_2}(\mathbf{v}_2)\big)\T \\
  \vdots  \\
 \big(\mathbf{v}_N-\mathpzc{P}_{\mathcal{A}_N}(\mathbf{v}_N)\big)\T
 \end{pmatrix}.
\end{align}

 Denote\footnote{In the rest of the proof we sometimes omit $K$ in $\mathbf{v}_i(K)$, $\mathbf{v}(K)$, and $\mathbf{v}_{\rm ave}(K)$ in order to simplify the presentation. One should however keep in mind that they always depend on $K$.} $\mathbf{v}_{\rm ave}(K)=\sum_{i=1}^N \mathbf{v}_i(K)$. Let
 \begin{align}
\mathcal{M}:= \Big\{\mathbf{w}=( \mathbf{w}_1\T \dots  \mathbf{w}_N\T)\T: \ \mathbf{w}_1=\dots=\mathbf{w}_N\Big\}.
\end{align}
be the consensus manifold. Denote $\lambda_2(\mathbf{L}^\S)>0$ as the second smallest eigenvalue of $\mathbf{L}^\S$. We can now conclude that
\begin{align}
( K \lambda_2(\mathbf{L}^\S) )^2\|\mathbf{v}\|_{\mathcal{M}}^2
 & \stackrel{a)}{\leq} \Big\|K \mathbf{L}^\S\otimes \mathbf{I}_m \mathbf{v}\Big\|^2 \nonumber\\
 &\stackrel{b)}{=} \sum_{i=1}^N \big\| \mathbf{v}_i-\mathpzc{P}_{\mathcal{A}_i}(\mathbf{v}_i) \big\|^2 \nonumber\\
  &\stackrel{c)}{\leq} N B_0^2
\end{align}
where  $a)$ holds from the fact that the zero eigenvalue of $\mathbf{L}^\S$ corresponds to a unique unit eigenvector whose entries  are all the same, $b)$ is from (\ref{32}), and $c)$ holds from the definition of $B_0$ and the fact that $\mathbf{v}\in\mathcal{Z}_K$. This allows us to further derive
\begin{align}
(K \lambda_2(\mathbf{L}^\S))^2 \|\mathbf{v}\|_{\mathcal{M}}^2  = (K \lambda_2(\mathbf{L}^\S))^2\sum_{i=1}^N \|\mathbf{v}_i -\mathbf{v}_{\rm ave}\|^2 \leq N B_0^2,
\end{align}
and then \begin{align}
\sum_{i=1}^N \|\mathbf{v}_i -\mathbf{v}_{\rm ave}\|^2 \leq \frac{ N B_0^2}{ (K \lambda_2(\mathbf{L}^\S))^2}.
\end{align}
Therefore, for any $\varsigma>0$ we can find $K_1(\varsigma)>0 $ that
\begin{align}\label{31}
\sum_{i=1}^N \big\|\mathbf{v}_i(K) -\mathbf{v}_{\rm ave}(K)\big\| \leq \varsigma,
\end{align}
for all $K\geq K_1(\varsigma)$.

\noindent {\bf Step 2}. Applying Lemma \ref{lem:affineprojection}, we have \begin{align}\label{52}
U(\mathbf{y}):=\|\mathbf{z}-\mathbf{H}\mathbf{y}\|^2= \sum_{i=1}^N \big|\mathbf{z}_i-\mathbf{h}_i\T \mathbf{y}\big|^2=\sum_{i=1}^N \big\|\mathbf{y}\big\|_{\mathcal{A}_i}^2.
\end{align}
Then with (\ref{31}) and noticing  the continuity of  $U(\cdot)$, for any $\varsigma>0$, there exists $\varrho>0$ such that
\begin{align}\label{51}
\Big|U( \mathbf{v}_i) - U(\mathbf{v}_{\rm ave})\Big|\leq \varsigma,\ i=1,\dots,N
\end{align}
if
\begin{align}\label{201}
\sum_{i=1}^N \big\|\mathbf{v}_i(K) -\mathbf{v}_{\rm ave}(K)\big\| \leq \varrho.
\end{align}
 Consequently, we can conclude without loss of generality\footnote{If $\varrho \geq \varsigma$ then we can replace $\varrho$ with $\varsigma$ in (\ref{201}) with (\ref{51}) continuing to hold; if $\varrho < \varsigma$ then both (\ref{31}) and (\ref{51}) hold directly. } that for any $\varsigma>0$ we can find $K_1(\varsigma)>0 $ so that both  (\ref{31}) and (\ref{51})  hold when  $K\geq K_1(\varsigma)$.

Now noticing $\mathbf{1}\T\mathbf{L}^\S=0$, from (\ref{32}) we have
\begin{align}
\sum_{i=1}^N\big(\mathbf{v}_i-\mathpzc{P}_{\mathcal{A}_i}(\mathbf{v}_i)\big)=0.
\end{align}
Therefore, with (\ref{31}), we can find another $K_2(\varsigma)$ such that
\begin{align}\label{53}
\Big\|\sum_{i=1}^N\big(\mathbf{v}_{\rm ave}-\mathpzc{P}_{\mathcal{A}_i}(\mathbf{v}_{\rm ave})\big)\Big\|\leq \varsigma/C_0
\end{align}
for all $K\geq K_2(\varsigma) $.

Take $K_\ast(\epsilon)=\max\{1,K_1(\epsilon/2),K_2(\epsilon/4) \}$. We can finally conclude that
\begin{align}
\Big|U(\mathbf{y}^\star) - U(\mathbf{v}_i)\Big|&\stackrel{a)}{\leq} \Big|U(\mathbf{y}^\star) - U(\mathbf{v}_{\rm ave})\Big|+\frac{\epsilon}{2} \nonumber\\
&\stackrel{b)}{\leq}\Big\| \nabla U(\mathbf{v}_{\rm ave})\Big\|\cdot \big\|\mathbf{y}^\star-\mathbf{v}_{\rm ave}\big\|+\frac{\epsilon}{2}\nonumber\\
&\stackrel{c)}{=}2\Big\| \sum_{i=1}^N\big(\mathbf{v}_{\rm ave}-\mathpzc{P}_{\mathcal{A}_i}(\mathbf{v}_{\rm ave})\big) \Big\|\cdot \big\|\mathbf{y}^\star-\mathbf{v}_{\rm ave}\big\|+\frac{\epsilon}{2}\nonumber\\
&\stackrel{d)}{\leq}\frac{\epsilon}{2C_0} \cdot \big\|\mathbf{y}^\star-\mathbf{v}_{\rm ave}\big\|+\frac{\epsilon}{2}\nonumber\\
&\stackrel{e)}{\leq}{\epsilon}
\end{align}
for all $i\in\mathrm{V}$, where $a)$ is from (\ref{51}), $b)$ is from the convexity of $U$, $c)$ is based on direct computation of $\nabla U$ in (\ref{52}), $d)$ is due to (\ref{53}), and $e)$ holds because $\big\|\mathbf{y}^\star-\mathbf{v}_{\rm ave}\big\|\leq C_0$ from the definition of $C_0$ and $\mathbf{v}_{\rm ave}$.

 This completes the proof of Theorem \ref{thm5}.

\subsection{Proof of Theorem \ref{thm6}}

Consider the following dynamics for the considered network model:
\begin{align}\label{60}
\dot{q}_i=K\sum\limits_{j \in \mathrm{N}_i(t)}a_{ij}(t)\big(q_j-q_i\big)+w_i(t), \ i\in \mathrm{V}
\end{align}
where $q_i\in\mathbb{R}$, $K>0$ is a given constant, $a_{ij}(t)$ are weight functions satisfying our standing assumption, and  $w_i(t)$ is a piecewise continuous  function. The proof is based on the following lemma on the robustness of  consensus subject to noises, which is a special case of Theorem 4.1 and Proposition 4.10  in \cite{shisiam}.

\begin{lemma}\label{lemrobust}
Let $\mathrm{G}_{\sigma(t)}$ be $\delta$-$\mathsf{UJSC}$ with respect to $T>0$. Then along (\ref{60}), there holds that for any $\epsilon>0$,  there exist a sufficiently small $T_\epsilon>0$  and sufficiently large $K_\epsilon$ such that  $$
\limsup_{t\rightarrow +\infty} \big |q_i(t)-q_j(t)\big|\leq  \epsilon\|w(t)\|_{\infty}
$$
for all initial value $q^0$ when $K\geq K_\epsilon$ and $T\leq T_\epsilon$, where $ \|w(t)\|_{\infty}:= \max_{i \in\mathrm{V}} \sup_{t\in[0,\infty)} |w_i(t)|$.
\end{lemma}

With Assumption [A1], the set
$$
\Delta_{\mathbf{x}(0)}:= \bigg[ {\rm co} \Big( \mcup_{i=1}^N\mathcal{W}(\mathbf{x}_i(0))\Big)\bigg]^N
$$
is a compact set, and is obviously positively invariant along the flow (\ref{1}). Therefore, we can define
\begin{align*}
&D_{\mathbf{x}(0)}:= \max_{i\in \mathrm{V}} \sup \Big\{ \big\|\mathpzc{P}_{\mathcal{A}_i}(\mathbf{u}_i)-\mathbf{u}_i\big\|:  \mathbf{u}=(\mathbf{u}_1\T \dots \mathbf{u}_N\T)\T\in\Delta_{\mathbf{x}(0)} \Big\}.
\end{align*}
as a finite number. Now along the trajectory $\mathbf{x}(t)$ of (\ref{1}), we have
$$
\big\|\mathpzc{P}_{\mathcal{A}_i}(\mathbf{x}_i(t))-\mathbf{x}_i(t)\big\|\leq D_{\mathbf{x}(0)}
$$
for all $t\geq 0$ and all $i\in\mathrm{V}$. Invoking Lemma \ref{lemrobust} we have  for any $\epsilon>0$ and  any initial value $\mathbf{x}(0)$, there exist $K_0(\epsilon,\mathbf{x}(0))>0$ and $T_0(\epsilon,\mathbf{x}(0))$ such that
 $$
\limsup_{t\to \infty} \big\|\mathbf{x}_i(t)-\mathbf{x}_j(t)\big\|\leq \epsilon, \ \forall i,j\in\mathrm{V}
$$
if $K\geq K_0$ and $T\leq T_0$.

Furthermore,  with Lemma \ref{lem:subspace-affine-projection}, we have
\begin{align}
&\frac{d}{dt}\Big( \mathpzc{P}_{\mathcal{A}_i}(\mathbf{x}_i)-\mathbf{x}_i\Big)=K\Big( \sum_{j\in \mathrm{N}_i(t)}a_{ij}(t)\big(\mathpzc{P}_{\mathcal{A}_i}(\mathbf{x}_j)-\mathpzc{P}_{\mathcal{A}_i}(\mathbf{x}_i)\big) \Big)+\Big(1+K \sum_{j\in \mathrm{N}_i(t)}a_{ij}(t)\Big) \mathpzc{P}_{\mathcal{A}_i}(0)\nonumber\\
 &- K\Big( \sum_{j\in \mathrm{N}_i(t)}a_{ij}(t)\big(\mathbf{x}_j-\mathbf{x}_i\big) \Big)- \Big(\mathpzc{P}_{\mathcal{A}_i}(\mathbf{x}_i)-\mathbf{x}_i\Big),
\end{align}
which implies
\begin{align}\label{66}
\frac{d}{dt}\sum_{i=1}^N\Big( \mathpzc{P}_{\mathcal{A}_i}(\mathbf{x}_i)-\mathbf{x}_i\Big)=-\sum_{i=1}^N\Big( \mathpzc{P}_{\mathcal{A}_i}(\mathbf{x}_i)-\mathbf{x}_i\Big)
\end{align}
if Assumption [A2] holds and $\mathrm{G}_{\sigma(t)}$ is balanced. While  if $\mathbf{x}(0)\in \mathcal{A}_1\times\cdots \times \mathcal{A}_N$, then $$
\sum_{i=1}^N\Big( \mathpzc{P}_{\mathcal{A}_i}(\mathbf{x}_i(0))-\mathbf{x}_i(0)\Big)=0.
$$
This certainly guarantees $\sum_{i=1}^N\Big( \mathpzc{P}_{\mathcal{A}_i}(\mathbf{x}_i(t))-\mathbf{x}_i(t)\Big)=0$ for all $t\geq 0$ in view of (\ref{66}). The proof for the fact that
 $$
\limsup_{t\to \infty} \big\|\mathbf{x}_i(t)-\mathbf{y}^\star\big\|\leq \epsilon, \ i\in\mathrm{V}
$$
 can then be built using exactly the same analysis as the final part of the proof of Theorem \ref{thm5}.

We have now completed the proof of Theorem \ref{thm6}.

\subsection{Proof of Theorem \ref{thm7}}
The proof is based on the following lemma.

\begin{lemma}
There holds $\mathpzc{P}_{\mathcal{A}_m}({\sum_{i=1}^N \mathbf{x}_i}/{N})=\sum_{i=1}^N\mathpzc{P}_{\mathcal{A}_m}({\mathbf{x}_i})/N$ for all $m\in\mathrm{V}$.
\end{lemma}
{\it Proof.} From Lemma \ref{lem:subspace-affine-projection} we can easily know $\mathpzc{P}_{\mathcal{K}} (\mathbf{y} +\mathbf{u})=\mathpzc{P}_{\mathcal{K}}(\mathbf{y})+\mathpzc{P}_{\mathcal{K}}(\mathbf{u})-\mathpzc{P}_{\mathcal{K}}(0)$ for all $\mathbf{y},\mathbf{u}\in\mathbb{R}^m$. As a result, we have
\begin{align}
\mathpzc{P}_{\mathcal{A}_m}({\sum_{i=1}^N \mathbf{x}_i})=N\mathpzc{P}_{\mathcal{A}_m}({\sum_{i=1}^N \mathbf{x}_i}/{N})- (N-1)\mathpzc{P}_{\mathcal{K}}(0).
\end{align}
On the other hand,
\begin{align}
\mathpzc{P}_{\mathcal{A}_m}({\sum_{i=1}^N \mathbf{x}_i})=\sum_{i=1}^N \mathpzc{P}_{\mathcal{A}_m}({\mathbf{x}_i})- (N-1)\mathpzc{P}_{\mathcal{K}}(0).
\end{align}
The desired lemma thus holds. \hfill$\square$

Suppose $\mathrm{G}_{\sigma(t)}=\mathrm{G}^\S$ is fixed, complete, and $a_{ij}(t) \equiv a^\S >0$ for all $i,j\in\mathrm{V}$. Now along the flow (\ref{2}), we have
\begin{align}
\frac{d}{dt}\frac{\sum_{i=1}^N \mathbf{x}_i}{N} &=\sum_{i=1}^N  \sum_{j=1}^Na^\S\big(\mathpzc{P}_{\mathcal{A}_i}(\mathbf{x}_j)-\mathpzc{P}_{\mathcal{A}_i}(\mathbf{x}_i)\big)/N \nonumber\\
&=\sum_{i=1}^N  \sum_{j=1}^Na^\S\big(\mathpzc{P}_{\mathcal{A}_i}(\mathbf{x}_j)-\mathbf{x}_i\big)/N \nonumber\\
&=a^\S  \sum_{i=1}^N \bigg[\frac{\sum_{j=1}^N \mathpzc{P}_{\mathcal{A}_i}(\mathbf{x}_j)}{N}   -  \frac{\sum_{j=1}^N \mathbf{x}_j} {N}\bigg] \nonumber\\
&=a^\S  \sum_{i=1}^N \bigg[\mathpzc{P}_{\mathcal{A}_i}\Big(\frac{ \sum_{j=1}^N\mathbf{x}_j}{N}\Big)   -\frac{\sum_{j=1}^N \mathbf{x}_j}{N}\bigg]
\end{align}
for any  initial value  $\mathbf{x}(0)\in \mathcal{A}_1\times\cdots \times \mathcal{A}_N$.  The desired result follows straightforwardly and this concludes the proof of Theorem \ref{thm7}.

\begin{figure*}
\begin{minipage}[t]{0.5\linewidth}
\centering
\includegraphics[width=3in]{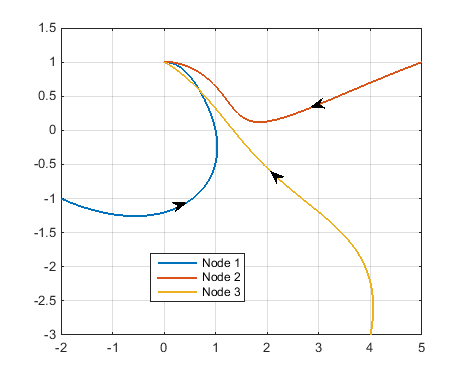}
\end{minipage}%
\begin{minipage}[t]{0.5\linewidth}
\centering
\includegraphics[width=3in]{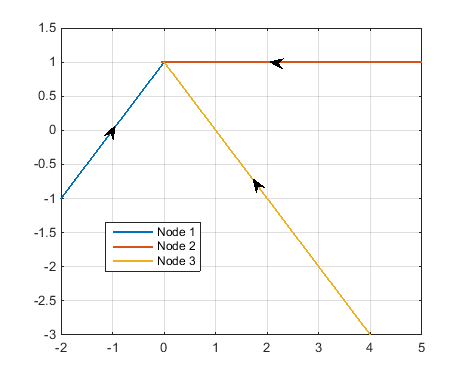}
\end{minipage}
\caption{The trajectories of the node states along the ``consensus + projection" flow (left) and the ``projection consensus" flow (right).}
\label{fig:fig1}
\end{figure*}

\begin{figure*}
\begin{minipage}[t]{0.5\linewidth}
\centering
\includegraphics[width=3in]{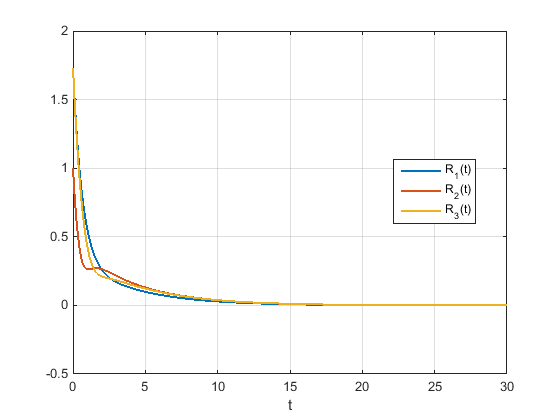}
\end{minipage}%
\begin{minipage}[t]{0.5\linewidth}
\centering
\includegraphics[width=3in]{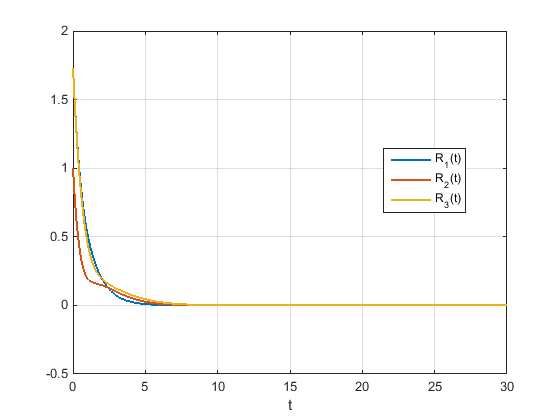}
\end{minipage}
\caption{The trajectories of  $R_i(t)$  along the ``consensus + projection" flow (left) and the ``projection consensus" flow (right).}
\label{fig:fig2}
\end{figure*}

\section{Numerical Examples}\label{Sec:numerical}
In this section, we provide a few examples illustrating the established results.

\medskip

\noindent {\bf Example 1}. Consider three nodes in the set $\{1,2,3\}$ whose interactions form a fixed, three-node directed cycle. Let $m=2$, $K=1$, and $a_{ij}=1$ for all $i,j\in\mathrm{V}$.  Take $\mathbf{h}_1 =(-1/\sqrt{2}\ 1/\sqrt{2})\T$, $\mathbf{h}_2 =(0\ 1)\T$, $\mathbf{h}_1 =(1/\sqrt{2}\ 1/\sqrt{2})\T$ and $z_1=1/\sqrt{2}$, $z_2=1$,  $z_3=1/\sqrt{2}$ so the linear equation (\ref{LinearEquation}) has a unique solution at $(0,1)$ corresponding to Case (I). With the same initial value $\mathbf{x}_1(0)=(-2\ -1)\T$, $\mathbf{x}_2(0)=(5\ 1)\T$, and $\mathbf{x}_3(0)=(4\ -3)\T$, we plot the trajectories of $\mathbf{x}(t)$, respectively, for the ``consensus + projection" flow (\ref{1}) and the ``projection consensus" flow (\ref{2}) in Figure \ref{fig:fig1}. The plot is consistent with the result of Theorems \ref{thm1} and \ref{thm2}.

\medskip

\noindent {\bf Example 2}. Consider three nodes in the set $\{1,2,3\}$ whose interactions form a fixed, three-node directed cycle.  Let $m=3$, $K=1$, and $a_{ij}=1$ for all $i,j\in\mathrm{V}$.   Take $\mathbf{h}_1 =(-1/\sqrt{2}\ 1/\sqrt{2}\ 0)\T$, $\mathbf{h}_2 =(0\ 1\ 0)\T$, $\mathbf{h}_1 =(1/\sqrt{2}\ 1/\sqrt{2}\ 0)\T$ and $z_1=1/\sqrt{2}$, $z_2=1$,  $z_3=1/\sqrt{2}$ so corresponding to Case (II), the linear equation (\ref{LinearEquation}) admits an infinite solution set  $$
\mathcal{A}:=\Bigg\{\mathbf{y}\in\mathbb{R}^3:  \begin{pmatrix}
  \mathbf{h}_1\T \\
 \mathbf{h}_2\T
 \end{pmatrix}\mathbf{y}=\begin{pmatrix}
  z_1\T \\
 z_2\T
 \end{pmatrix}\Bigg\}.
$$
 For the initial value  $\mathbf{x}_1(0)=(1\ 2\  3 )\T$, $\mathbf{x}_2(0)=(  -1\ 1\ 2)\T$, and $\mathbf{x}_3(0)=(1\ 0\  1)\T$, we plot the trajectories of $$
 R_i(t):=\Big\|\mathbf{x}_i(t)-\sum_{i=1}^3 \mathpzc{P}_\mathcal{A} (\mathbf{x}_i(0))/3\Big\|, \ i=1,2,3
 $$
 respectively, for the ``consensus + projection" flow (\ref{1}) and the ``projection consensus" flow (\ref{2})  in Figure \ref{fig:fig2}. The plot is consistent with the result of Theorems \ref{thm3}, \ref{thm4}, and \ref{prop1}.

\medskip

\begin{figure}
\centering
\includegraphics[width=3.2in]{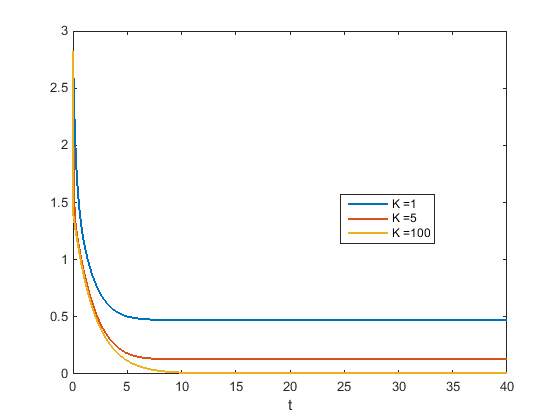}
\caption{The evolution of $ E_K(t)$  for $K=1,5,100$, respectively along the ``consensus + projection" flow.}
\label{fig:fig3}
\end{figure}

\noindent {\bf Example 3}. Consider four nodes in the set $\{1,2,3,4\}$ whose interactions form a fixed, four-node undirected cycle.  Let $m=2$,  and $a_{ij}=1$ for all $i,j\in\mathrm{V}$.   Take $\mathbf{h}_1 =(-1/\sqrt{2}\ 1/\sqrt{2})\T$, $\mathbf{h}_2 =(1/\sqrt{2}\ 1/\sqrt{2})\T$, $\mathbf{h}_3=(-1/\sqrt{2}\ 1/\sqrt{2})\T $, $\mathbf{h}_4=(1/\sqrt{2}\ 1/\sqrt{2})\T $ and $z_1=1/\sqrt{2}$,   $z_3=1/\sqrt{2}$, $z_3=-1/\sqrt{2}$, $z_3=-1/\sqrt{2}$ so corresponding to Case (III) the linear equation (\ref{LinearEquation}) has  a unique least squares solution $\mathbf{y}^\star=(0\ 0)$.
 For the initial value $\mathbf{x}_1(0)=(0\ 1)\T$, $\mathbf{x}_2(0)=(1\ 0)\T$, $\mathbf{x}_3(0)=(2\ 1)\T$, and $\mathbf{x}_4(0)=(-1\ 0)\T$ we plot the trajectories of $$
 E_K(t):=\sum_{i=1}^4\Big\|\mathbf{x}_i(t)-\mathbf{y}^\star\Big\|^2,
 $$
along  the ``consensus + projection" flow (\ref{1}) for $K=1, 5, 100$, respectively in Figure \ref{fig:fig3}.   The plot is consistent with the result of Theorem \ref{thm5}.

\section{Conclusions}\label{Sec:conclusions}

Two distributed network flows were studied  as distributed solvers for a linear algebraic equation $\mathbf{z}=\mathbf{H}\mathbf{y}$, where a  node $i$ holds a row $\mathbf{h}_i\T$ of the matrix $\mathbf{H}$ and the entry $z_i$ in the vector $\mathbf{z}$. A ``consensus + projection" flow consists of two terms, one from standard consensus dynamics and the other as projection onto each affine subspace specified by the $\mathbf{h}_i$ and $z_i$.  Another  ``projection consensus" flow  simply replaces the relative state feedback in consensus dynamics with projected relative state feedback. Under mild graph conditions, it was shown that  that all node states converge to a common solution of the linear algebraic equation, if there is any. The convergence is global for the ``consensus + projection" flow while local for the ``projection consensus" flow.  When the linear equation has no exact solutions but has a well defined least squares approximate solution, it was proved  that the node states can converge to somewhere  arbitrarily near the least squares solution as long as the gain of the consensus dynamics is sufficient large for ``consensus + projection" flow under fixed and undirected graphs.  It was also shown that the ``projection consensus" flow drives the average of the node states to the least squares solution if the communication graph is complete. Numerical examples were provided verifying  the established results.  Interesting future direction includes more precise   comparisons of the existing continuous-time or discrete-time linear equation solvers in terms of convergence speed, computational complexity, and communication complexity.

\section*{Appendix}

\subsection*{A. Proof of Lemma \ref{lem:directed}. (i)}
Suppose $h_\S>0$. We show   $\lim_{t\to\infty}\|\mathbf{x}_i(t)\|_{\mathcal{A}^\sharp}=h_\S$ for all $i\geq \mathrm{V}$ by a contradiction argument. Suppose (to obtain a contradiction) that there exists a node  $i_0\in\mathrm{V}$  with $l_\S:=\liminf_{t\to\infty} \|\mathbf{x}_{i_0}(t)\|_{\mathcal{A}^\sharp}<h_\S$. Therefore, we can find a time instant $t_\ast>T_{\epsilon}$ with
\begin{align}
 \|\mathbf{x}_{i_0}(t_\ast)\|_{\mathcal{A}^\sharp}\leq \sqrt{\frac{h_\S^2+l_\S^2}{2}}.
\end{align}
In other words, there is an absolute positive distance between $\|\mathbf{x}_{i_0}(t_\ast)\|_{\mathcal{A}^\sharp}$ and the limit $h_\S$ of $ \max_{\in\mathrm{V}} \|\mathbf{x}_i(t)\|_{\mathcal{A}^\sharp}$.

\noindent  Let  $\mathrm{G}_{\sigma(t)}$ be $\delta$-$\mathsf{UJSC}$. Consider the $N-1$ time intervals $[t_\ast,t_\ast +T],\cdots,[t_\ast+(N-2)T,t_\ast +(N-1)T]$. In view of the arguments in (\ref{8}), we similarly obtain
\begin{align}\label{12}
\frac{d}{dt} \big\|\mathbf{x}_{i_0}(t)\big\|_{\mathcal{A}^\sharp}^2&=2\Big\langle \mathbf{x}_{i_0}(t)-\mathpzc{P}_{\mathcal{A}^\sharp}(\mathbf{x}_{i_0}(t)), 
 \sum_{j\in \mathrm{N}_{i_0}(t)}a_{i_0j}(t)\big(\mathpzc{P}_{i_0}(\mathbf{x}_j)-\mathpzc{P}_{i_0}(\mathbf{x}_{i_0})\big)\Big\rangle\nonumber\\
&{\leq}   \sum_{j\in \mathrm{N}_{i_0}(t)}{a_{i_0j}(t)} \Big(\big\|\mathbf{x}_j(t)\big\|_{\mathcal{A}^\sharp}^2 - \big\|\mathbf{x}_{i_0}(t)\big\|_{\mathcal{A}^\sharp}^2  \Big),
\end{align}
which leads to
\begin{align}
\frac{d}{dt}\big \|\mathbf{x}_{i_0}(t)\big\|_{\mathcal{A}^\sharp}^2
&{\leq}   \sum_{j\in \mathrm{N}_{i_0}(t)}{a_{i_0j}(t)} \Big( (h_\S +\epsilon)^2 - \big\|\mathbf{x}_{i_0}(t)\big\|_{\mathcal{A}^\sharp}^2  \Big)
\end{align}
for all $ t\geq T_\epsilon$ noticing (\ref{10}). Denoting $t^\ast:=t_\ast +(N-1)T$ and applying  Gr\"{o}nwall's inequality we can further conclude that
\begin{align}
&\big \|\mathbf{x}_{i_0}(t)\big\|_{\mathcal{A}^\sharp}^2
 \nonumber\\
 &\leq e^{-\int_{t_\ast}^t \sum_{j\in \mathrm{N}_{i_0}(s)}{a_{i_0j}(s)}ds } \big\|\mathbf{x}_{i_0}(t_\ast)\big\|_{\mathcal{A}^\sharp}^2+\Big(1- e^{-\int_{t_\ast}^t \sum_{j\in \mathrm{N}_{i_0}(s)}{a_{i_0j}(s)}ds }\Big) (h_\S +\epsilon)^2 \nonumber\\
&\leq e^{-\int_{t_\ast}^{t^\ast} \sum_{j\in \mathrm{N}_{i_0}(s)}{a_{i_0j}(s)}ds } \big\|\mathbf{x}_{i_0}(t_\ast)\big\|_{\mathcal{A}^\sharp}^2 +\Big(1- e^{-\int_{t_\ast}^{t^\ast}  \sum_{j\in \mathrm{N}_{i_0}(s)}{a_{i_0j}(s)}ds }\Big) (h_\S +\epsilon)^2 \nonumber\\
&\leq  \mu \big\|\mathbf{x}_{i_0}(t_\ast)\big\|_{\mathcal{A}^\sharp}^2+\big(1-\mu\big) (h_\S +\epsilon)^2 \nonumber\\
&\leq  \frac{\mu}{2} \cdot l_\S^2+\big(1-\frac{\mu}{2}\big) \cdot(h_\S +\epsilon)^2
\end{align}
for all $t\in[t_\ast,t^\ast]$, where $\mu=e^{- (N-1)T a^\ast } $.

Now that   $\mathrm{G}_{\sigma(t)}$ is $\delta$-$\mathsf{UJSC}$, there must be a node $i_1\neq i_0$ for which $(i_0,i_1)$ is a $\delta$-arc over the time  interval $[t_\ast,t_\ast+T)$. Thus, there holds
\begin{align}\label{11}
\frac{d}{dt} \big\|\mathbf{x}_{i_1}(t)\big\|_{\mathcal{A}^\sharp}^2
&{\leq}   \sum_{j\in \mathrm{N}_{i_1}(t)}{a_{i_1j}(t)} \Big(\big\|\mathbf{x}_j(t)\big\|_{\mathcal{A}^\sharp}^2 - \big\|\mathbf{x}_{i_1}(t)\big\|_{\mathcal{A}^\sharp}^2  \Big)\nonumber\\
&= \mathbb{I}_{(i_0,i_1)\in\mathrm{E}_{\sigma(t)}}{a_{i_1i_0}(t)} \Big(\big\|\mathbf{x}_{i_0}(t)\big\|_{\mathcal{A}^\sharp}^2 - \big\|\mathbf{x}_{i_1}(t)\big\|_{\mathcal{A}^\sharp}^2  \Big)+\sum_{j\in \mathrm{N}_{i_1}(t)\setminus\{i_0\}}{a_{i_1j}(t)} \Big(\big\|\mathbf{x}_j(t)\big\|_{\mathcal{A}^\sharp}^2 - \big\|\mathbf{x}_{i_1}(t)\big\|_{\mathcal{A}^\sharp}^2  \Big)\nonumber\\
&\leq  \mathbb{I}_{(i_0,i_1)\in\mathrm{E}_{\sigma(t)}}{a_{i_1i_0}(t)} \Big( \frac{\mu}{2} \cdot l_\S^2+\big(1-\frac{\mu}{2}\big) \cdot(h_\S +\epsilon)^2 - \big\|\mathbf{x}_{i_1}(t)\big\|_{\mathcal{A}^\sharp}^2  \Big) \nonumber\\
&\ \ \ +\sum_{j\in \mathrm{N}_{i_1}(t)\setminus\{i_0\}}{a_{i_1j}(t)} \Big((h_\S +\epsilon)^2- \big\|\mathbf{x}_{i_1}(t)\big\|_{\mathcal{A}^\sharp}^2  \Big)
\end{align}
for $t\in[t_\ast,t_\ast+T]$. Noticing the definition of $\delta$-arcs and that $ \|\mathbf{x}_{i_1}(t_\ast)\|_{\mathcal{A}^\sharp}^2\leq (h_\S+\epsilon)^2$, we invoke the Gr\"{o}nwall's inequality again and conclude from (\ref{11}) that
\begin{align}
\big \|\mathbf{x}_{i_1}(t_\ast+T)\big\|_{\mathcal{A}^\sharp}^2
 &\leq \frac{\mu l_\S^2}{2} \bigg[e^{-\int_{t_\ast}^{t_\ast+T} \sum_{j\in \mathrm{N}_{i_1}(s)}{a_{i_1j}(s)}ds } \int_{t_\ast}^{t_\ast+T}e^{\int_{t_\ast}^{t}  f_1(s)ds } f_1(t) dt\bigg]\nonumber\\
&  +\frac{\mu (h_\S+\epsilon)^2}{2} \bigg[1-e^{-\int_{t_\ast}^{t_\ast+T} \sum_{j\in \mathrm{N}_{i_1}(s)}{a_{i_1j}(s)}ds }\cdot\int_{t_\ast}^{t_\ast+T}e^{\int_{t_\ast}^{t}  f_1(s)ds } f_1(t) dt\bigg]\nonumber\\
&\leq  \frac{\mu \gamma}{2} \cdot l_\S^2+\big(1-\frac{\mu\gamma}{2}\big) \cdot(h_\S +\epsilon)^2
\end{align}
 where $ f_1(t):=\mathbb{I}_{(i_0,i_1)\in\mathrm{E}_{\sigma(t)}}a_{i_1i_0}(t)$ and $\gamma=e^{- T a^\ast }(1-e^{-\delta}) $. This further allows us to apply the estimation of $\|\mathbf{x}_{i_0}(t_\ast+T)\|_{\mathcal{A}^\sharp}^2 $ over the interval $[t_\ast,t^\ast]$ to node $i_1$ for the interval $[t_\ast+T,t^\ast]$ and obtain
 \begin{align}\label{15}
\big \|\mathbf{x}_{i_1}(t_\ast+T)\big\|_{\mathcal{A}^\sharp}^2
&\leq  \frac{\mu^2 \gamma}{2} \cdot l_\S^2+\big(1-\frac{\mu^2\gamma}{2}\big) \cdot(h_\S +\epsilon)^2
\end{align}
for all $t\in[t_\ast+T,t^\ast]$.
Since  $\mathrm{G}_{\sigma(t)}$ is $\delta$-$\mathsf{UJSC}$, the above analysis can be recursively applied to the intervals $[t_\ast+T,t_\ast+2T),\dots,[t_\ast+(N-2)T,t_\ast+(N-1)T)$, for which nodes $i_2,\dots,i_{N-1}$ can be found, respectively, with $\mathrm{V}=\{i_0,\dots,i_{N-1}\}$ such that
 \begin{align}
\big \|\mathbf{x}_{i_m}(t^\ast)\big\|_{\mathcal{A}^\sharp}^2
\leq  \frac{\mu^{N-1} \gamma}{2} \cdot l_\S^2+\big(1-\frac{\mu^{N-1}\gamma}{2}\big) \cdot(h_\S +\epsilon)^2,
\end{align}
for $m=0,1,\dots,N-1.$
This implies
 \begin{align}\label{13}
h^\sharp(t^\ast)
&\leq  \frac{\mu^{N-1} \gamma}{2} \cdot l_\S^2/2+\big(1-\frac{\mu^{N-1}\gamma}{2}\big) \cdot(h_\S +\epsilon)^2 /2\nonumber\\
&<h_\S^2/2
\end{align}
when $\epsilon$ is sufficiently small. However, we have known that non-increasingly there holds $
\lim_{t\to \infty}h^\sharp(t)=h_\S^2/2$, and therefore such $i_0$ does not exist, i.e.,  $\lim_{t\to\infty}\|\mathbf{x}_i(t)\|_{\mathcal{A}^\sharp}=h_\S$ for all $i\geq \mathrm{V}$. The statement of Lemma \ref{lem:directed}. (i) is proved.

\subsection*{B. Proof of Lemma \ref{lem:directed}. (ii)}

Suppose $\lim_{t\to\infty}\|\mathbf{x}_i(t)\|_{\mathcal{A}^\sharp}=h_\S$ for all $i\geq \mathrm{V}$. Then for any $\epsilon>0$, there exists $\hat{T}_{\epsilon}>0$ such that
\begin{align}\label{rr10}
h_\S-\epsilon\leq \big\|\mathbf{x}_i(t)\big\|_{\mathcal{A}^\sharp}\leq h_\S+\epsilon,\ \forall t\geq  \hat{T}_{\epsilon}.
\end{align}
Moreover, for the ease of presentation  we assume that $\mathbf{h}_1,\dots,\mathbf{h}_N$ are distinct vectors since otherwise we can always combine the nodes with the same $\mathbf{h}_i$ as a cluster to be treated together in the following arguments.
Denote the angle between the two unit vectors $\mathbf{h}_{i}\neq \mathbf{h}_{j}$ as $\beta_{ij}\neq 0$. Then\footnote{Again, note that $\mathbf{x}_{j_\ast}(t)\in \mathcal{A}_{j_\ast}$ for all $t\geq 0$.}
\begin{align}
\Big\|\Big(I-\mathbf{h}_{i}\mathbf{h}_{i}^T\Big)\Big(\mathbf{x}_{j}-\mathpzc{P}_{\mathcal{A}^\sharp} \big(\mathbf{x}_{j} \big)\Big)\Big\|=\big|\cos (\beta_{ij})\big|\cdot \big \|\mathbf{x}_{j}\big\|_{\mathcal{A}^\sharp}.
\end{align}
This leads to (cf., (\ref{8}))
\begin{align}\label{19}
\frac{d}{dt} \big \|\mathbf{x}_{i}\big\|_{\mathcal{A}^\sharp}^2
&= 2\Big\langle \mathbf{x}_{i}(t)-\mathpzc{P}_{\mathcal{A}^\sharp}( \mathbf{x}_i(t)), \sum_{j\in \mathrm{N}_{i_\ast}(t)}a_{ij}(t)\big(\mathpzc{P}_{\mathcal{A}_i}(\mathbf{x}_j)-\mathpzc{P}_{\mathcal{A}_i}(\mathbf{x}_i)\big)\Big\rangle\nonumber\\
&\leq  \sum_{j\in \mathrm{N}_i(t)}{a_{ij}(t)} \Big(\Big\|\Big(I-\mathbf{h}_i\mathbf{h}_i^T\Big)\Big(\mathbf{x}_j(t)- \mathpzc{P}_{\mathcal{A}^\sharp}(\mathbf{x}_j(t))\Big)\Big\|^2 - \Big\|\Big(I-\mathbf{h}_i\mathbf{h}_i^T\Big)\Big(\mathbf{x}_i(t)-\mathbf{y}^\sharp\Big)\Big\|^2  \Big) \nonumber\\
&\leq\sum_{j\in \mathrm{N}_i(t)}{a_{ij}(t)} \Big(\big|\cos (\beta_{ij})\big|^2  \|\mathbf{x}_{j}\big\|_{\mathcal{A}^\sharp}^2 - \|\mathbf{x}_{i}\big\|_{\mathcal{A}^\sharp}^2 \Big) \nonumber\\
&\leq   \sum_{j\in \mathrm{N}_i(t)}{a_{ij}(t)}  \Big(\chi_\ast \|\mathbf{x}_{j}\big\|_{\mathcal{A}^\sharp}^2 - \|\mathbf{x}_{i}\big\|_{\mathcal{A}^\sharp}^2 \Big)\nonumber\\
&=  \sum_{j\in \mathrm{N}_i(t)}{a_{ij}(t)} \chi_\ast  \Big(\|\mathbf{x}_{j}\big\|_{\mathcal{A}^\sharp}^2 - \|\mathbf{x}_{i}\big\|_{\mathcal{A}^\sharp}^2 \Big) -(1-\chi_\ast)  \sum_{j\in \mathrm{N}_i(t)}{a_{ij}(t)} \|\mathbf{x}_{i}\big\|_{\mathcal{A}^\sharp}^2
\end{align}
with $\chi_\ast=\max_{i,j\in\mathrm{V}}\big|\cos (\beta_{ij})\big|^2<1$
for all $t\geq 0$.
This will in turn give us
\begin{align}\label{16}
&\frac{d}{dt} \big \|\mathbf{x}_{i}(t)\big\|_{\mathcal{A}^\sharp}^2
\leq 2\chi_\ast \epsilon \sum_{j\in \mathrm{N}_i(t)}{a_{ij}(t)} -(1-\chi_\ast)  \sum_{j\in \mathrm{N}_i(t)}{a_{ij}(t)} \|\mathbf{x}_{i}(t)\big\|_{\mathcal{A}^\sharp}^2
\end{align}
for all $t\geq \hat{T}_\epsilon$. Now we have
\begin{align}\label{17}
\int_{\hat{T}_\epsilon}^\infty  \sum_{j\in \mathrm{N}_i(t)}{a_{ij}(t)}dt=\infty
\end{align}
if  $\mathrm{G}_{\sigma(t)}$ is $\delta$-$\mathsf{UJSC}$. Combining (\ref{16}) and (\ref{17}) we arrive at
\begin{align}
 \limsup_{t\to\infty}\big \|\mathbf{x}_{i}(t)\big\|_{\mathcal{A}^\sharp}^2\leq \frac{2\chi_\ast}{1-\chi_\ast} \epsilon,
\end{align}
which leaves $h_\S=0$ the only possibility since $\epsilon$ can be arbitrary number. This proves Lemma \ref{lem:directed}. (ii).

\subsection*{C. Proof of Lemma \ref{lemma-bidirectional}}
 Again without loss of generality we can assume that  $\mathbf{h}_1,\dots,\mathbf{h}_N$ are distinct vectors. With (\ref{19}), we have
\begin{align}\label{19}
\frac{d}{dt} \sum_{i=1}^N \big \|\mathbf{x}_{i}\big\|_{\mathcal{A}^\sharp}^2&\leq  \sum_{i=1}^N\sum_{j\in \mathrm{N}_i(t)}{a_{ij}(t)} \chi_\ast  \Big(\|\mathbf{x}_{j}\big\|_{\mathcal{A}^\sharp}^2 - \|\mathbf{x}_{i}\big\|_{\mathcal{A}^\sharp}^2 \Big)\nonumber\\
 &-(1-\chi_\ast)  \sum_{i=1}^N\sum_{j\in \mathrm{N}_i(t)}{a_{ij}(t)} \|\mathbf{x}_{i}\big\|_{\mathcal{A}^\sharp}^2\nonumber\\
&=-(1-\chi_\ast)  \sum_{i=1}^N b_i(t) \|\mathbf{x}_{i}\big\|_{\mathcal{A}^\sharp}^2,
\end{align}
where $b_i(t):=\sum_{j\in \mathrm{N}_i(t)}{a_{ij}(t)}$. It is easy  to see from (\ref{19}) using  a contradiction argument that if  $\mathrm{G}_{\sigma(t)}$ is $\delta$-$\mathsf{BIJC}$, there must hold
 $$
 \lim_{t\to\infty} \sum_{i=1}^N \big \|\mathbf{x}_{i}\big\|_{\mathcal{A}^\sharp}^2 =0.
 $$
 Therefore, we conclude that $h_\S=0$ immediately and this proves the desired lemma.

\section*{Acknowledgement}
The authors  thank Zhiyong Sun for his generous help in the preparation of the numerical examples.

\end{document}